\documentclass[times,twocolumn,final]{elsarticle}

\usepackage{framed, multirow, multicol}

\usepackage{amssymb}
\usepackage{latexsym}

\usepackage{url}
\usepackage{xcolor}
\usepackage{hyperref}
\usepackage{epstopdf}
\usepackage{subfigure}
\usepackage{caption}
\usepackage{amsmath,graphicx}

\definecolor{newcolor}{rgb}{.8,.349,.1}

\begin{document}


\begin{frontmatter}

\title{SF2Former: Amyotrophic Lateral Sclerosis Identification From Multi-center MRI Data Using Spatial and Frequency Fusion Transformer}%



\author[inst1]{Rafsanjany Kushol\corref{cor1}}
\cortext[cor1]{Corresponding author: 
   email: kushol@ualberta.ca;  }
\author[inst2]{Collin C. Luk}
\author[inst2]{Avyarthana Dey}
\author[inst21]{Michael Benatar} 
\author[inst22]{Hannah Briemberg} 
\author[inst23]{Annie Dionne}
\author[inst24]{Nicolas Dupré} 
\author[inst25]{Richard Frayne} 
\author[inst26]{Angela Genge}
\author[inst27]{Summer Gibson}
\author[inst28]{Simon J. Graham}
\author[inst29]{Lawrence Korngut} 
\author[inst30]{Peter Seres}
\author[inst31]{Robert C. Welsh}
\author[inst30]{Alan Wilman} 
\author[inst32]{Lorne Zinman} 
\author[inst1,inst2]{Sanjay Kalra}
\author[inst1]{Yee-Hong Yang}

\address[inst1]{Department of Computing Science, University of Alberta, Edmonton,AB,Canada}
            
\address[inst2]{Department of Medicine, University of Alberta, Edmonton, AB, Canada}
            
\address[inst21]{Department of Neurology, University of Miami, Miami, FL, United States}
            
\address[inst22]{Department of Medicine, University of British Columbia, Vancouver, BC, Canada} 
            
\address[inst23]{CHU de Québec, Université Laval, Quebec, QC, Canada}
            
\address[inst24]{Department of Medicine, Université Laval, Quebec, QC, Canada}
 
\address[inst25]{Department of Radiology, University of Calgary, Calgary, AB, Canada} 
            
\address[inst26]{Department of Neurology and Neurosurgery, McGill University, Montreal, QC, Canada} 

\address[inst27]{Department of Neurology, University of Utah, Salt Lake City, UT, United States}

\address[inst28]{Department of Medical Biophysics, University of Toronto, Toronto, ON, Canada}
            
\address[inst29]{Department of Clinical Neurosciences, University of Calgary, 
Calgary, AB, Canada}            

\address[inst30]{Department of Biomedical Engineering, University of Alberta, 
Edmonton, AB, Canada}

\address[inst31]{Department of Psychiatry, University of Utah, Salt Lake City,
UT, United States}
            
\address[inst32]{Department of Medicine, University of Toronto, Toronto, ON, Canada}


\begin{abstract}
Amyotrophic Lateral Sclerosis (ALS) is a complex neurodegenerative disorder involving motor neuron degeneration. Significant research has begun to establish brain magnetic resonance imaging (MRI) as a potential biomarker to diagnose and monitor the state of the disease. Deep learning has turned into a prominent class of machine learning programs in computer vision and has been successfully employed to solve diverse medical image analysis tasks. However, deep learning-based methods applied to neuroimaging have not achieved superior performance in ALS patients classification from healthy controls due to having insignificant structural changes correlated with pathological features. Therefore, the critical challenge in deep models is to determine useful discriminative features with limited training data. By exploiting the long-range relationship of image features, this study introduces a framework named $SF^2Former$ that leverages vision transformer architecture's power to distinguish the ALS subjects from the control group. To further improve the network's performance, spatial and frequency domain information are combined because MRI scans are captured in the frequency domain before being converted to the spatial domain. The proposed framework is trained with a set of consecutive coronal 2D slices, which uses the pre-trained weights on ImageNet by leveraging transfer learning. Finally, a majority voting scheme has been employed to those coronal slices of a particular subject to produce the final classification decision. Our proposed architecture has been thoroughly assessed with multi-modal neuroimaging data (i.e., T1-weighted, R2*, FLAIR) using two well-organized versions of the Canadian ALS Neuroimaging Consortium (CALSNIC) multi-center datasets. The experimental results demonstrate the superiority of our proposed strategy in terms of classification accuracy compared with several popular deep learning-based techniques.
\end{abstract}


\end{frontmatter}


\section{Introduction}
\label{sec:intro}
ALS is a neurodegenerative disorder with an average age of onset in the late fifties and early sixties affecting the upper and lower motor neurons of the nervous system. Degeneration of upper motor neuron (UMN) is responsible for spasticity, exaggerated reflexes, and modest weakness, while degeneration of lower motor neuron (LMN) causes muscle atrophy, fasciculations, and more severe weakness. Patients may lose limb function and have trouble walking, speaking, and eventually breathing as the disease progresses. Respiratory failure is often the cause of death, and the average survival time from symptom onset is $3-5$ years. The pathophysiology underlying neurodegeneration in ALS is not adequately understood. While some cases of ALS are familial $(5-10)\%$, in the vast majority of patients, the cases are sporadic $(90-95)\%$. A paucity of pharmacologic therapies have been approved (i.e., riluzole and edaravone) for use in the early stage of the disease to slow disease progression and improve survival \cite{jaiswal2019riluzole}. However, there are no therapies that can halt disease progression. One reason behind this lack of treatments is the absence of established biomarkers that can follow disease progression or aid in early diagnosis. This is critical for conducting effective clinical trials on new therapies. Recent research has sought to establish MRI techniques as a biomarker of disease progression in ALS.

\begin{figure}[bth!]

\begin{minipage}[b]{1.0\linewidth}
  \centering
  \centerline{\includegraphics[width=\textwidth]{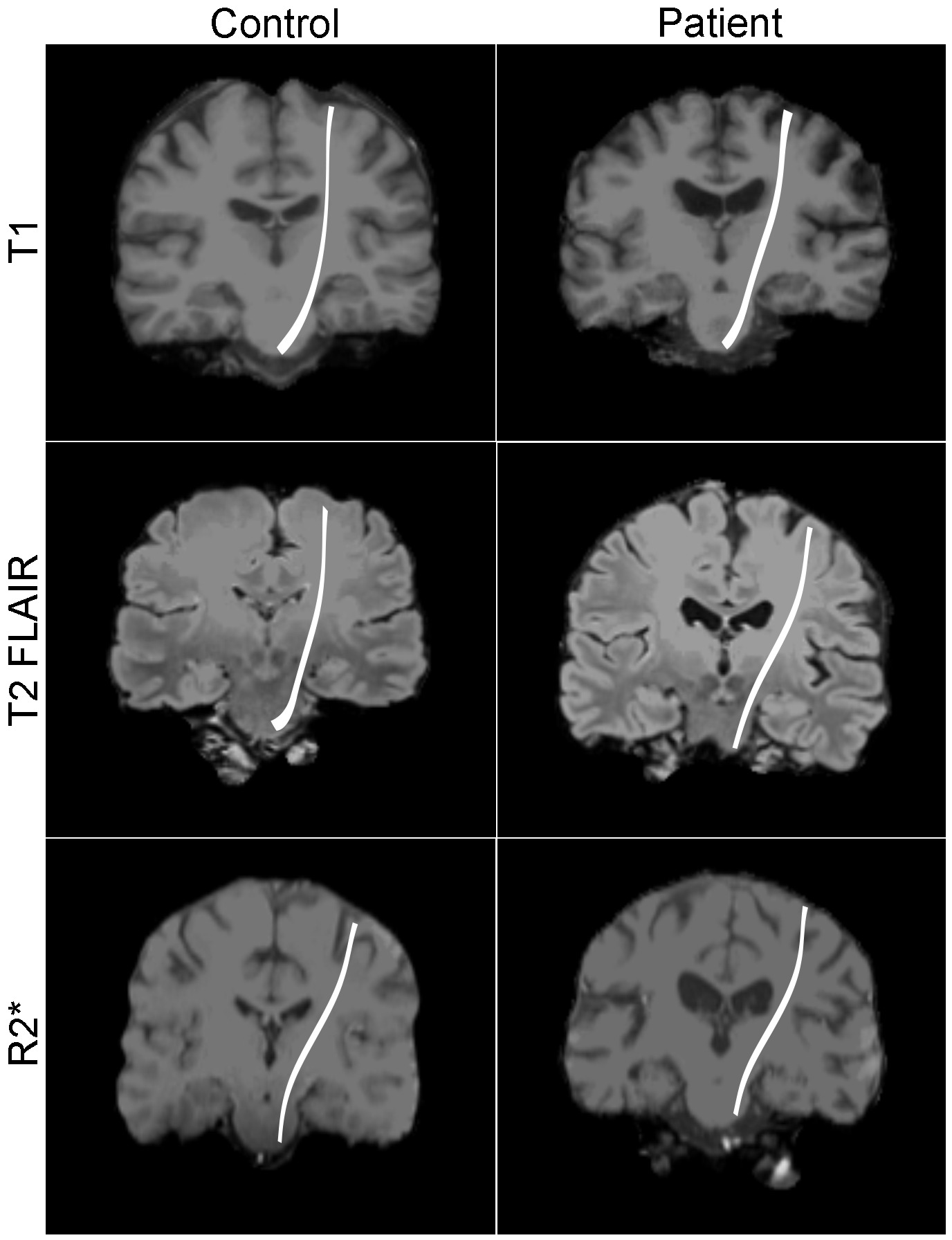}}
\end{minipage}
\caption{MR images of controls and ALS patients for T1-weighted, T2-FLAIR and R2* modalities. Coronal images were sampled at the plane of the precentral gyrus with a white line demonstrating the approximate path of the corticospinal tract within each plane. There are no visually discernible features in the gray and white matter between controls and ALS patients.}
\label{fig:intro}
\end{figure}

Magnetic resonance (MR) images are considered an essential diagnostic tool for many neurological disorders. Among various modalities of MRI, structural MRI, such as T1-weighted and T2-weighted images, are regularly utilized as a part of clinical investigation as they can illustrate the appearance and integrity of gray and white matter. FLAIR is an inversion recovery structural MRI sequence with a long inversion duration that suppresses the signal from cerebrospinal fluid (CSF) in the resultant scan. On the other hand, R2* (1/T2*) MRI can estimate iron concentration by non-invasive means. However, the brain regions affected by most neurological illnesses can be minimal, posing a challenge in the use of MRI for disease diagnosis. In the case of ALS, neuroimaging is routinely used to rule out differential diagnoses but has no reliable role in confirming diagnosis. In a minority of cases, visual inspection of affected regions such as the precentral gyrus (PCG) and corticospinal tract (CST) may demonstrate imaging changes \cite{sage2007quantitative, maani2016cerebral}, but this is not the case for the majority of scans of ALS patients (Figure \ref{fig:intro}). Hence, we sought to utilize a deep learning-based procedure to automatically distinguish MRI brain scans of ALS patients from healthy controls. 

Deep learning-based algorithms have recently dominated many research fields. For example, convolutional neural networks (CNNs) have become the most prevailing framework for automatic medical image processing applications such as disease classification \cite{wen2020convolutional, playout2022focused}, tissue segmentation \cite{pinaya2022unsupervised, kushol2020rbvs}, image registration \cite{chen2022transmorph} etc. The transformer architecture, developed by Vaswani et al. \cite{vaswani2017attention}, is the most widely used paradigm in natural language processing (NLP) domain. Inspired by the success of the self-attention-based deep neural networks of transformer models in NLP, Dosovitskiy et al. \cite{dosovitskiy2020image} presented the vision transformer (ViT) architecture for the image classification task. Despite the popularity of CNNs in image processing applications, ViT models have demonstrated higher performance in multiple image analysis contexts. This is because CNN models perform poorly in learning long-range information due to their confined receptive fields, limiting their capacity for vision tasks. On the other hand, frequency domain features are another essential context in image processing tasks that is barely investigated in transformer-based deep models correlated to medical image analysis. This study leverages the potential of ViT models, which utilizes both spatial and frequency domain features to achieve a satisfactory classification performance in ALS disease.

This research introduces an effective and robust transformer-based framework titled $SF^2Former$ (spatial and frequency fusion transformer), that is subsequently used to classify ALS subjects from healthy controls in multi-modal brain imaging data. The summary of our noteworthy contributions is listed as follows:\\
1) Adopting the concept of the ViT to distinguish ALS samples from healthy controls by employing a set of intermediate 2D coronal slices of 3D MRI volume.\\
2) Linear fusion of spatial and frequency domain information in a simple yet efficient way to better obtain robust local and global discriminative features.\\
3) Implementing a majority voting procedure on chosen coronal images of the identical subject to produce the final classification outcome enhances the overall accuracy. At the same time, it also automates testing with any unknown sample instantly from a 3D MRI input scan.\\
4) To the best of our knowledge, this is the first transformer-based deep model study for ALS classification that achieves state-of-the-art performance compared with many popular CNN-based deep learning methods.

\section{Literature Review}
\label{sec:literature} 

\subsection{ALS diagnosis}
\label{ssec:als_dis_diag}

The involvement of iron accumulation in the motor cortex area has been reported in multiple in vivo and ex vivo studies for ALS cohorts. Compared to T1-weighted, T2-weighted and FLAIR, the T2*-weighted sequence can better capture hypointensities in ALS patients, which is prominent in precentral gyrus gray matter (PGGM) \cite{ignjatovic2013brain}. Wang et al. \cite{wang2020methods} found that ALS subjects have increased R2* response in the primary motor cortex compared to healthy controls. Hecht et al. \cite{hecht2001mri} reveal more frequent hyperintense signals at the CST in FLAIR scans than T1-weighted, T2-weighted and proton density-weighted images. Similar findings are also demonstrated by Jin et al. \cite{jin2016hyperintensity} with higher CST hyperintensity for the ALS subjects compared to the control samples in the subcortical PCG. Fabes et al. \cite{fabes2017quantitative} show that FLAIR intensity is significant in the CST and the corpus callosum in the ALS group compared to that of normal controls.

Liu et al. \cite{liu2021voxelhop} propose a model named VoxelHop using T2-weighted structural MR images to detect ALS. However, they evaluated in a small-scale dataset composed of 20 controls and 26 patients. By utilizing recurrent neural networks and random forest classifiers, Thome et al. \cite{thome2022classification} design a feature set from structural and functional resting-state MRI. Nevertheless, the best classification accuracy they end up with is $66\%$ after analyzing various combinations of feature sets. On the other hand, Elahi et al. \cite{elahi2020texture} introduce a modified co-occurrence histogram of oriented gradients (M-CoHOG) method for feature selection using 2D coronal slices of T1-weighted images. This technique achieves $76\%$ classification accuracy in the single-center dataset but results in poor consistency in an extended version of the multicenter database. On top of that, it requires laborious effort from experts to manually select the appropriate coronal slices for each individual. Moreover, Chen et al. \cite{chen2020identification} employ fractional anisotropy (FA) information of diffusion tensor image (DTI) and linear kernel support vector machine (SVM) to classify ALS vs. healthy controls and obtains $83\%$ classification accuracy. However, their dataset is also limited, comprising 22 ALS patients and 26 healthy subjects. In another study, using DTI and texture analysis with linear SVM classifier authors report $80\%$ sensitivity and specificity \cite{kocar2021multiparametric}.  

\subsection{Transformer in medical image analysis}
\label{ssec:trans}
The transformer's architecture \cite{vaswani2017attention} was first introduced in the context of NLP. It allows for the capture of long-term dependencies as well as the processing of several words or patches in parallel. The application of transformers in computer vision is limited due to the high computational cost. To minimize the spatial dimension of the representation, the ViT \cite{dosovitskiy2020image} embeds an image into non-overlapping patch tokens. On the famous computer vision ImageNet dataset \cite{deng2009imagenet}, the ViT provides a new state-of-the-art performance for image classification. One of the flaws of the ViT is that it is incapable of learning the dependency within the patch. The Swin transformer \cite{liu2021Swin} leverages the relationship from local to global using a hierarchical structure. Moreover, the global filter network (GFNet) \cite{rao2021global} has been proposed for capturing both long-term and short-term spatial relationships in the Fourier domain. By applying a discrete Fourier transform with a global convolution, the GFNet reconstructs ViT's self-attention layer, considerably improving performance. Touvron et al. \cite{touvron2021training} introduces data-efficient image transformers (DeiT) using knowledge distillation that allow ViT to perform well on smaller datasets as well.\\
Transformer-based approaches are not only leading in computer vision tasks but have also successfully been applied to diverse medical image analysis contexts. TransUNet \cite{chen2021transunet} uses CNNs to extract features, which are then fed into a ViT network for efficient medical image segmentation tasks. TransFuse \cite{zhang2021transfuse} also leverages ViT and CNNs by fusing their features for various 2D and 3D medical image segmentation. In contrast, MedT \cite{valanarasu2021medical}, which is based on axial-attention, investigates the viability of using transformers without large-scale datasets. Mok and Chung \cite{mok2022affine} developed coarse to fine vision transformer (C2FViT) for 3D affine medical image registration using ViT and a multi-resolution strategy. Utilizing the effectiveness of ViT, ScoreNet \cite{stegmuller2022scorenet} has been proposed for histopathological image classification, whereas Uni4Eye \cite{cai2022uni4eye} has been developed for the robust ophthalmic image classification task. SphereMorph \cite{cheng2020cortical} is a diffeomorphic cortical surface registration network which uses a UNet-style architecture and a modified spatial transformer layer.The success of these models demonstrates the transformer's enormous promise in medical image analysis.

\section{Proposed Method}
\label{sec:proposed}
\begin{figure}[htb]

\begin{minipage}[b]{1.0\linewidth}
  \centering
  \centerline{\includegraphics[width=\textwidth]{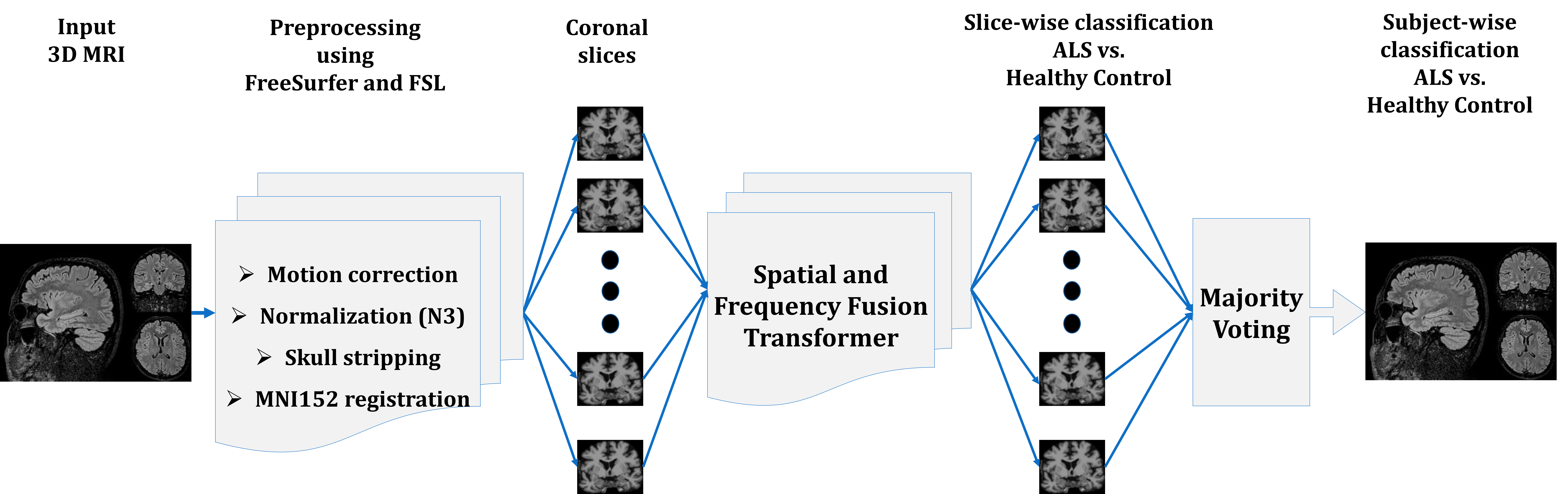}}
\end{minipage}
\caption{The overall workflow of the proposed stages.}
\label{fig:proposed}
\end{figure}

The proposed framework includes simple preprocessing steps of the raw data using FreeSurfer \cite{fischl2012freesurfer} and FSL \cite{jenkinson2012fsl},  selecting a fixed range of 2D coronal slices, combining the features of two transformer networks in the spatial and the Fourier domain, and majority voting on the predictions of individual slices to finalize the classification result. Figure \ref{fig:proposed} depicts the overall workflow of our proposed phases.

\subsection{Preprocessing}
\label{ssec:preprocessing}
A straightforward, fast, and easy to perform preprocessing pipeline has been followed on the original 3D brain MRI to make data suitable for the deep models. We execute a number of common preprocessing operations with the FreeSurfer program, such as motion correction, skull stripping and non-parametric non-uniform intensity normalization (N3). (Command: \textit{recon-all -subject subjectname -i input-file.nii -autorecon1}). Then we perform registration to MNI-152 standard-space using FSL \textit{flirt} function. After the reconstruction of the original images, the new shape becomes $182\times218\times182$, and the voxel dimension converts to $1\times1\times1$  $mm^3$. We use an eight-core CPU platform that leverages parallel processing and takes around 5 minutes per scan on average to perform the preprocessing. Otherwise, processing each subject individually without parallelization takes approximately 15 minutes per sample.

\subsection{Slice selection}
\label{ssec:slice}
\begin{figure}[htb]

\begin{minipage}[b]{1.0\linewidth}
  \centering
  \centerline{\includegraphics[width=\textwidth]{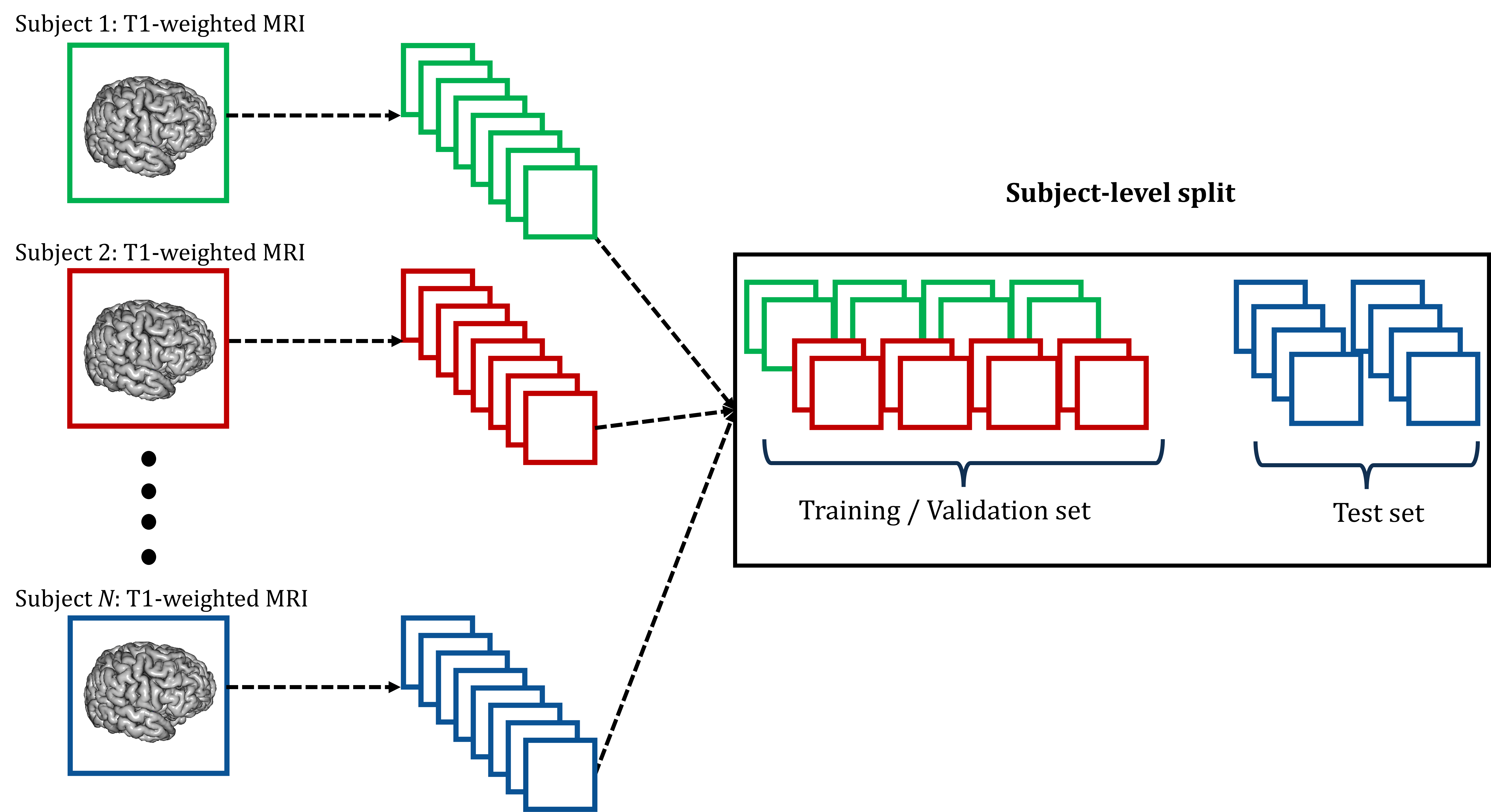}}
\end{minipage}
\caption{Subject-level split process for the data to train our proposed model.}
\label{fig:subject-level-split}
\end{figure}

After conducting empirical analysis, we have found that the top performance is accomplished by manipulating the coronal slices among the three different planes (coronal, sagittal, and axial) of 3D MRI scans. After that, we explore a wide range of slice combinations by training and testing the network to figure out a potential zone of meaningful slices. The detailed outcome of our experiments with various clusters of slices is given in section \ref{ssec:slice_selection}. Eventually, 15 consecutive 2D images from the central section of the coronal plane are used to train the proposed framework. Manual observation from experts suggests that this range of coronal slices better captures the CST, which is a prominent region of interest in ALS. It is essential to mention that the slices generated from the same subject will never be used in both train or test sets. In other words, our method follows subject-level split protocol to avoid data leakage, which is further illustrated in Fig \ref{fig:subject-level-split}. Data leakage in a machine learning model refers to the fact that the information from the test and training data sets is mistakenly shared. As a result, the model is already familiar with some aspects of the test data after training. Indeed, a recent study \cite{yagis2021effect} found that many prior neurodegenerative disease classification approaches did not follow a proper division of slices in their training or testing data, and hence, their results convey incorrect and overly optimistic classification accuracies. 

\subsection{Spatial and frequency fusion transformer}
\label{ssec:fusion_trans}
\begin{figure*}[htb]
    \begin{minipage}[b]{1.0\linewidth}
    \centering
    \centerline{\includegraphics[width=\textwidth]{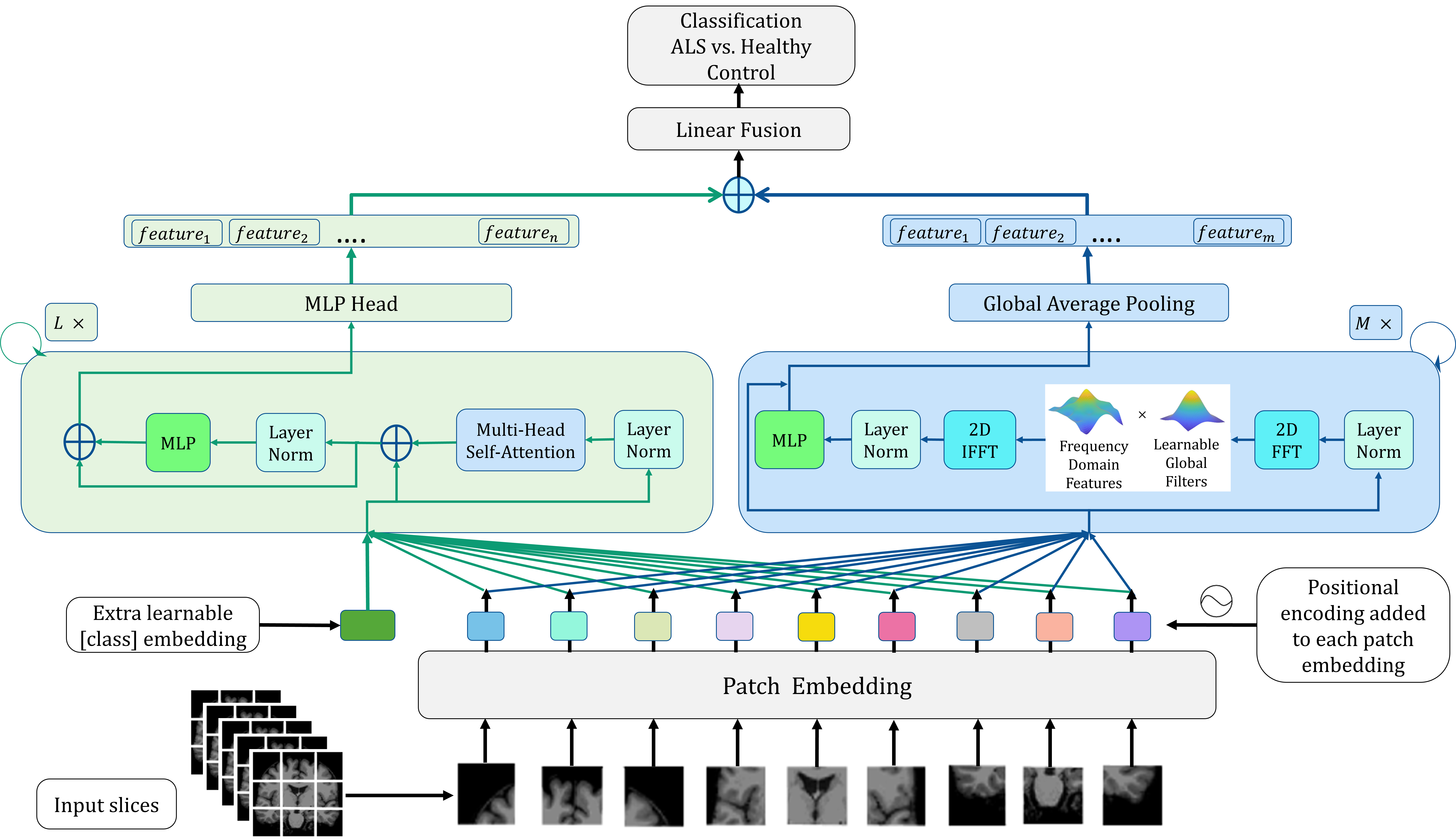}}
    \end{minipage}
\caption{Proposed $SF^2Former$ architecture. The left branch of the methodology encodes features from the spatial domain, whereas the right segment encodes features from the frequency domain. Finally, the linear fusion module incorporates the features to assemble the classification decision for each 2D slice.}
\label{fig:fusion}
\end{figure*}

Figure \ref{fig:fusion} depicts the overall architecture of our proposed $SF^2Former$ method, which integrates features from two vision transformer-based networks. One network is responsible for generating features from the spatial domain, and the other is capable of developing features from the frequency domain.
 
The ViT is the first successful method by adopting the transformer that has achieved state-of-the-art performance in many computer vision tasks. Unlike other deep learning-based practices that extract features from images using convolution, which is challenging in capturing long-term dependencies, ViT employs self-attention to solve this problem. However, because matrix multiplication has a high computational cost in self-attention, it embeds the image in patch tokens and uses patch tokens as input to lower the computational complexity. The left branch of our proposed architecture, similar to ViT, consists of alternating layers of multiheaded self-attention (MSA) and an MLP block with two layers of Gaussian Error Linear Unit (GELU) \cite{hendrycks2016gaussian} non-linearity on top of the encoder. Before each block, layer norm (LN) is applied, and residual connections are adjusted after each block. $L$ represents the number of transformer encoder layers which is 12 in our case.

To begin with, we have 2D slices with a spatial resolution of $(H, W)$ and $C$ channels. As an input to the transformer, we reshape the image $x \hspace{1mm} \in \hspace{1mm} \mathbb{R}^{H\times W \times C}$ into a series of 2D patches $x_p \hspace{1mm} \in \hspace{1mm} \mathbb{R}^{N\times (P^2.C)}$. Here $(P, P)$ is the size of each patch, and we  have $N = HW/P^2$ number of total patches, which is the input sequence length for the transformer. Now, we flatten the patches and project it to $D$ dimensions with a trainable linear projection for the reason that the transformer maintains a constant latent vector size $D$ throughout all of its layers. The output of this mapping is referred to as patch embeddings which is  shown in Fig \ref{fig:fusion}. To preserve positional information, position embeddings $\mathit{E_{pos}} \hspace{1mm} \in \hspace{1mm} \mathbb{R}^{(N+1)\times D}$ are appended to the patch embeddings. We also use typical learnable 1D position embeddings, as indicated in a previous work \cite{dosovitskiy2020image}. We add a learnable extra class embedding to the series of embedded patches $z^0_0 = x_{class}$, the state of which functions as the image representation $y = LN(z^0_L)$ at the transformer encoder's output. The value of $z$ at different layers and positions can be expressed as follows:
\begin{align}
    \mathit{z_0} &= [x_{class}; x^1_p\mathit{E}; x^2_p\mathit{E};\cdots; x^N_p\mathit{E}] + \mathit{E_{pos}} \hspace{10mm} \mathit{E} \hspace{1mm} \in \hspace{1mm} \mathbb{R}^{(P^2.C)\times D} \\
 \mathit{z'_l} &= MSA(LN(z_{l-1})) + z_{l-1} \hspace{30mm} l=1\cdots L\\
\mathit{z_l} &= MLP(LN(\mathit{z'_l})) +\mathit{z'_l} \hspace{37mm} l=1\cdots L
\end{align}

There is a close relationship between MRI data acquisition and the frequency domain. The mathematical formation of raw MRI details is initially performed in the frequency domain, then converted to the spatial domain to make it interpretable visually. Hence, it motivates us also to extract features in the frequency domain. However, directly involving a deep model in the frequency domain is computationally expensive, especially when the image resolution is high. The  GFNet~\cite{rao2021global} uses a similar downsampling design of the ViT and replaces the self-attention layer with the Fast Fourier Transform (FFT) on the embedded patch tokens. The right side branch of our proposed architecture introduced in Fig \ref{fig:fusion} basically follows the concept of the GFNet. The essential objective of this network is to learn the frequency domain interactions among different spatial positions. Another notable difference of the GFNet compared to the ViT is the global average pooling in the final feature map as an alternative to the extra class embedding head. The GFNnet accepts non-overlapping $H \times W$ patches as input and flattens them into $L = HW$ tokens with a dimension of $D$. Each spatial domain token $x \hspace{1mm} \in \hspace{1mm} \mathbb{R}^{H\times W \times D}$ transformed by 2D FFT  generates a complex tensor $\mathit{X}$ in the frequency domain as:
\begin{equation}
    \mathit{X} = 2D \hspace{1mm} \mathit{FFT} [x] \hspace{1mm} \in \hspace{1mm} \mathbb{C}^{H\times W \times D}.
\end{equation}

Then, we modulate $\mathit{X}$ (the spectrum of $x$) with respect to a learnable filter $\mathit{K}$ in the form of element-wise multiplication and can be expressed as:
\begin{equation}
    \tilde{\mathit{X}} = \mathit{X} \hspace{1mm} \odot \hspace{1mm} \mathit{K}.
\end{equation}

The parameter $\mathit{K}$ is a global filter representing an arbitrary frequency-domain filter with the same dimension as $\mathit{X}$. Ultimately, the modulated spectrum $\tilde{\mathit{X}}$ is transformed back to the spatial domain using the inverse FFT (iFFT), and the tokens are updated as:

\begin{equation}
    x = 2D \hspace{1mm} \mathit{iFFT} \hspace{1mm} [\tilde{\mathit{X}}].
\end{equation}
 
The other components, e.g. layer norm and MLP, used in the diagram for the GFNet are identical to that in the ViT. Among different variants of the GFNet, we adopt the transformer-style GFNet-B version, which consists of 19 layers/depth and an embedding dimension of 512. So the values of $m$ and $M$ becomes 512 and 19, respectively, in our presented diagram of Fig. \ref{fig:fusion}. To take advantage of information from both the spatial and frequency domain, we propose a new linear fusion block to combine the features extracted from the ViT and the GFNet. We construct a new linear head equal to the joint embedding dimensions of the ViT and the GFNet, where the input comes from the final layers of these two networks as a form of concatenation. The output from the linear fusion block containing the merged features of the spatial and frequency domain is used to carry out the classification decision. The loss function we operate throughout our model is the cross-entropy (CE) loss function.

\subsection{Majority voting}
\label{ssec:majority}
The science behind applying majority voting in our methodology is that it is unlikely to present all the disease-affected tissues or areas in all the chosen slices. In other words, it is not feasible to automatically capture the 2D planes that will always contain distinct clinical characteristics for any subjects without manual effort from a specialist. To alleviate the false-positive response from those insignificant slices, we leverage the idea of majority voting at the end of the architecture. The final classification of an individual sample will be based on the maximum number of the same class identified in a given span of slices. The ablation study section also reports the effectiveness of this majority voting scheme in enhancing classification accuracy.

\section{Experimental Analysis}
\label{sec:experimental}

\subsection{Dataset}
\label{ssec:dataset}
Neuroimaging data is obtained from two independent datasets of the Canadian ALS Neuroimaging Consortium (CALSNIC\footnote{\url{(https://calsnic.org/)}}) \cite{kalra2020canadian} study. The CALSNIC is a multi-center and multi-modal longitudinal study where 3T MRI scans are collected from three different scanner manufacturers (i.e., GE, Philips, and Siemens). The data used in our experimentation with CALSNIC1 is accumulated from five centers (i.e., Calgary, Edmonton, Toronto, Vancouver, and Montreal), whereas CALSNIC2 data comprises seven different centers (i.e., Calgary, Edmonton, Toronto, Quebec, Miami, Utah, and Montreal). To avoid potential data leakage issue between training and testing, we only consider MRI data of participants with a certain visit (baseline) in our experiments. Due to a shortage of data in CALSNIC1, the FLAIR and R2* modalities are only considered from the CALSNIC2 dataset. The overview of the demographics of the datasets is illustrated in table \ref{tab:demographic}.

\begin{table*}[h]
    \footnotesize
	\centering
	\caption{Demographic detail of T1-weighted MR images for the CALSNIC1 and CALSNIC2 datasets}
	\label{tab:demographic}
	\begin{tabular}{c|ccc|ccc}
		
		\hline
		 {\textbf {Participant }} & \multicolumn{3}{c}{\textbf {CALSNIC1}} & \multicolumn{3}{|c}{\textbf {CALSNIC2}}\\
		 \textbf{characteristics} & {\textbf {ALS }} &  {\textbf {Healthy}} & \textbf{\multirow{2}{*} {p-value}} & {\textbf {ALS }} &  {\textbf {Healthy}} & \textbf{\multirow{2}{*} {p-value}}\\
	     & \textbf{patients} & \textbf{controls} & & \textbf{patients} & \textbf{controls} & \\
		\hline
		
		Subjects & 61 & 59 & - & 116 & 116 & - \\
		\hline
		      Sex: Male/Female  & 36/25 & 27/32 & 0.15 & 76/40 & 57/59 & 0.01*\\
        \hline
		      Age (years)  &  &  & &  &  &\\
		      Mean $\pm$ S.D. & 58.4 $\pm$ 10.7 & 54.0 $\pm$ 10.2 & 0.02* & 60.1 $\pm$ 10.1 & 56.3 $\pm$ 10.6 & 0.005*\\
		      Median & 57.0 & 55.0 & - & 60.9 & 58.7 & -\\
		      Range & 33.0 - 86.0 & 25.0 - 69.0 & - & 25.6 - 83.4 & 25.8 - 77.0 & -\\
        \hline
		      ALSFRS-R score  &  &  & &  &  &\\
		      Mean $\pm$ S.D. & 39.2 $\pm$ 5.0 & - & - & 37.3 $\pm$ 7.0 & - & -\\
		      Median & 40.0 & - & - & 39.0 & - & -\\
		      Range & 22.0 - 47.0 & - & - & 7.0 - 47.0 & - & -\\
        \hline
		      Symptom duration  &  &  & &  &  &\\
		      (months)  &  &  & &  &  &\\
		      Mean $\pm$ S.D. & 16.1 $\pm$ 10.5 & - & - & 22.6 $\pm$ 13.2 & - & -\\
		      Median & 13.3 & - & - & 19.3 & - & -\\
		      Range & 4.0 - 54.8 & - & - & 2.6 - 59.8 & - & -\\
	
		\hline

		\hline
		
	\end{tabular}
\end{table*}

\subsection{Implementation}
\label{ssec:implementation}
The proposed framework is implemented using PyTorch \cite{paszke2019pytorch} and runs on a server with 4 NVIDIA 2080 Ti GPUs. Our coding follows the publicly available implementation of the ViT\footnote{\url{(https://github.com/jeonsworld/ViT-pytorch)}} and the GFNet\footnote{\url{(https://github.com/raoyongming/GFNet)}}. Data augmentation with random rotations and flipping is used to prepare a robust training batch. We resize each coronal slice to $224 \times 224$. After performing normalization, the intensity of pixel values is in the range of [0,1]. We employ the SGD optimizer to train all the networks with a momentum value of 0.9. We choose an initial learning rate of 0.001 and decay the rate to $10^{-5}$ using the cosine schedule. We use a patch size of $16 \times 16$, a total number of 150 epochs, and a batch size of 16 for both transformer networks. The final accuracy and the values of hyper-parameters reported in the study are attained from five-fold cross-validation (CV). Specifically, we design a variation of the Stratified KFold approach where each fold is conditioned on a similar share of data from all the available centers. It ensures that each fold will incorporate approximately the same percentage of their respective center's samples. For a better understanding of data distribution for model training of CALSNIC1 and CALSNIC2, figure \ref{fig:cross_val} is presented below. The data split ratio of train: validation: test data is 7:1:2 in each fold. The training time of our proposed approach takes approximately 6 hours operating on a single GPU of 12GB memory.

\begin{figure}[htb]

\begin{minipage}[b]{1.0\linewidth}
  \centering
  \centerline{\includegraphics[width=\textwidth]{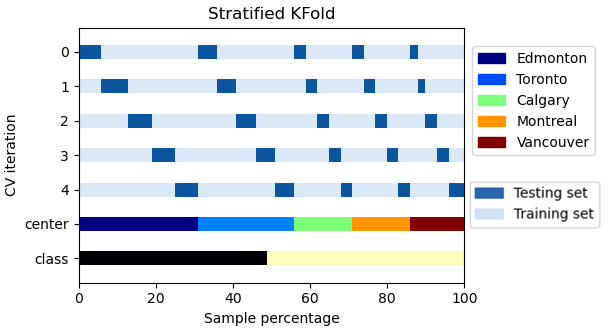}}
  \centerline{(a) CALSNIC1}\medskip
\end{minipage}
\hfill
\begin{minipage}[b]{1.0\linewidth}
  \centering
  \centerline{\includegraphics[width=\textwidth]{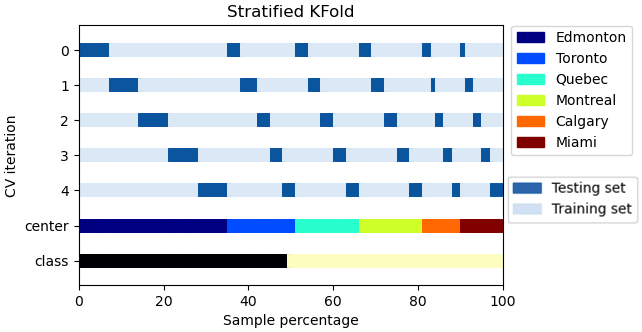}}
  \centerline{(b) CALSNIC2}\medskip
\end{minipage}
\caption{Stratified five-fold cross-validation (CV) designed for CALSNIC datasets. The row labelled  `class' indicates the percentage of ALS patients and healthy controls, the number of which is similar and balanced in both datasets. Next, the row tagged `center' shows the percentage of participants in the corresponding dataset from available centers. The five rows above the line `center' show training and test set distribution with five iterations of CV. Each iteration involves a similar proportion of samples from each center.}
\label{fig:cross_val}
\end{figure}

\subsection{Results}
\label{ssec:results}
\begin{figure*}[htb]

\begin{minipage}[b]{.46\linewidth}
  \centering
  \centerline{\includegraphics[width=\textwidth]{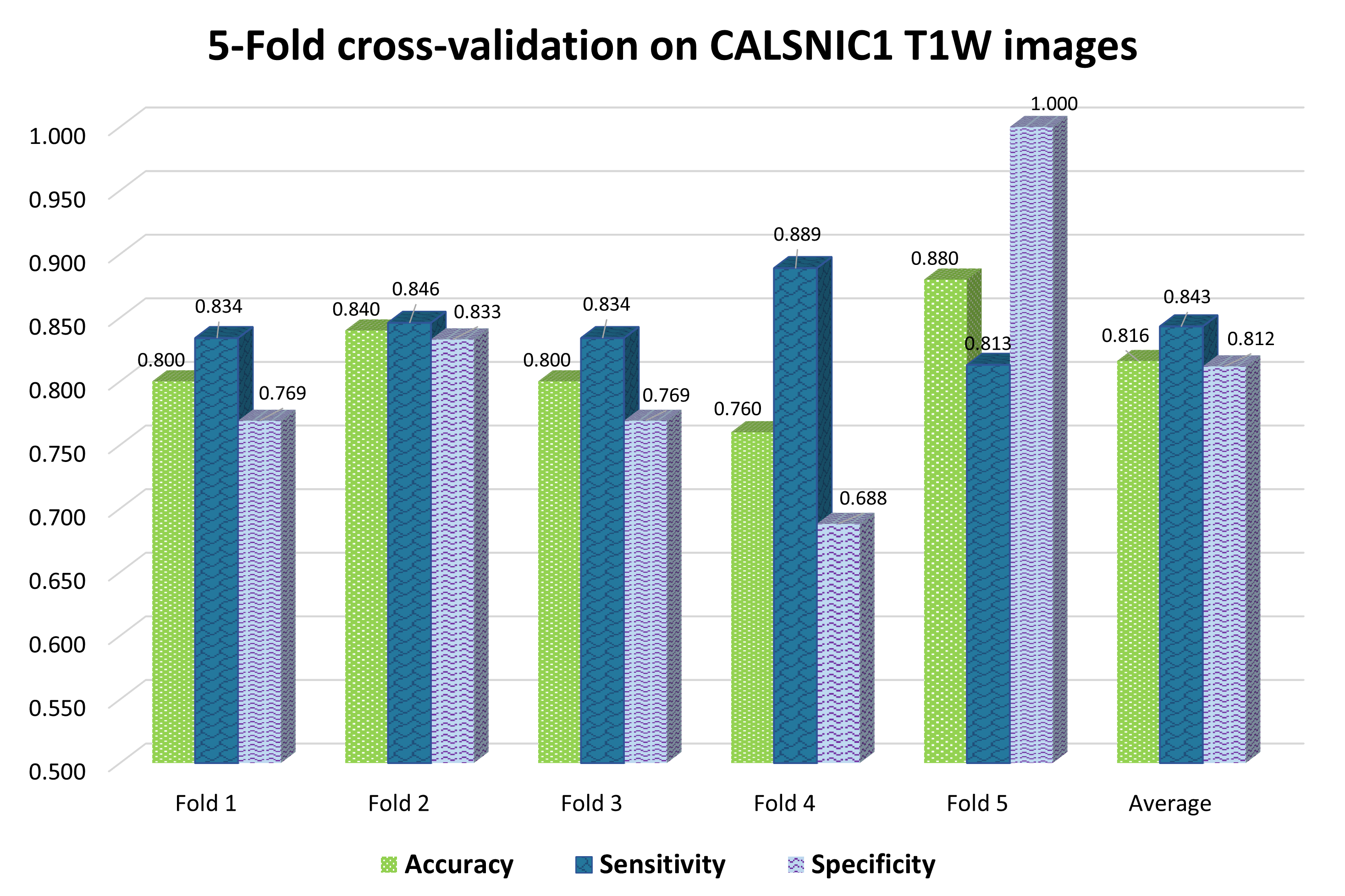}}
  \centerline{(a) CALSNIC1 T1W}\medskip
\end{minipage}
\hfill
\begin{minipage}[b]{0.46\linewidth}
  \centering
  \centerline{\includegraphics[width=\textwidth]{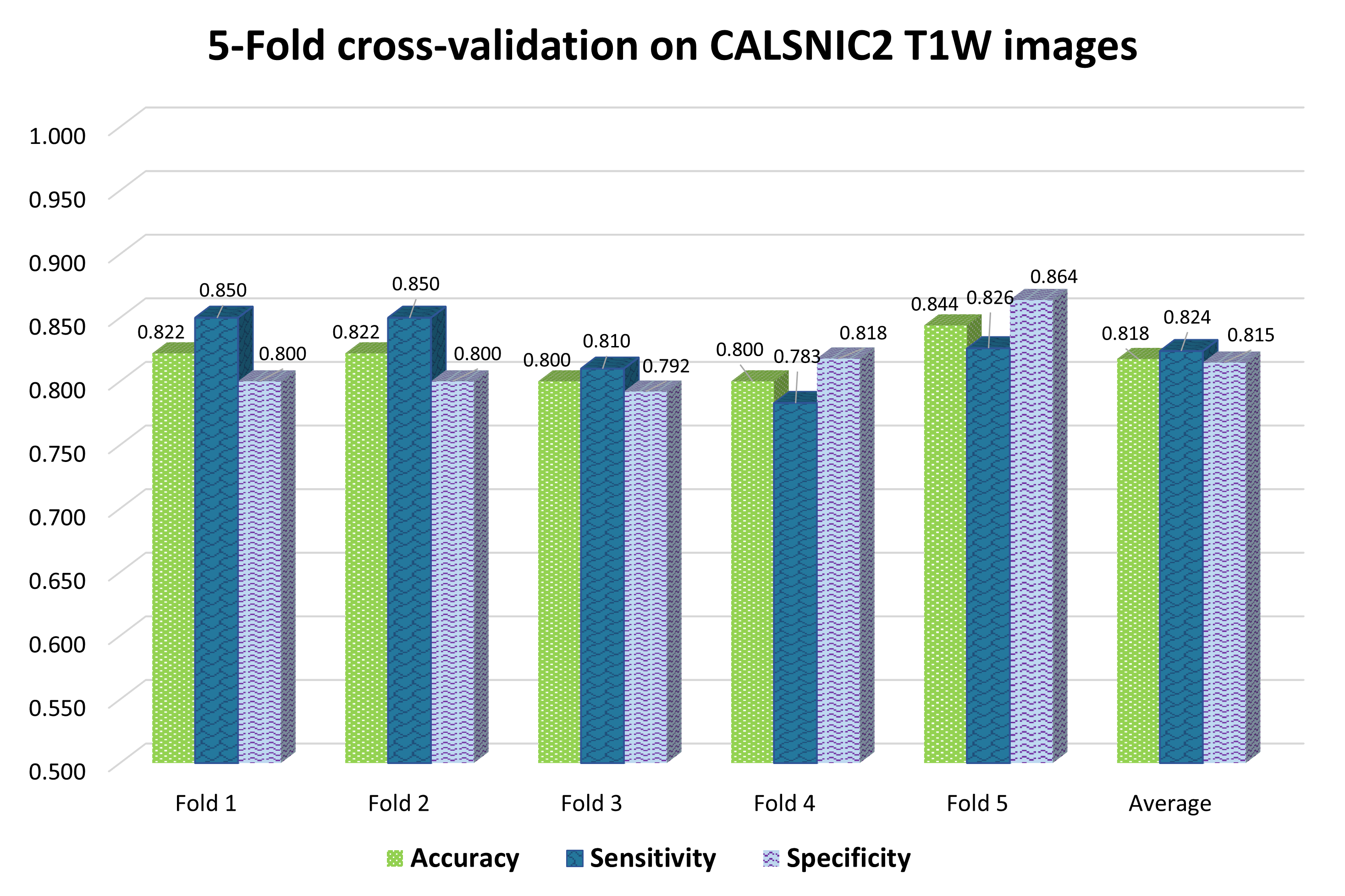}}
  \centerline{(b) CALSNIC2 T1W}\medskip
\end{minipage}

\begin{minipage}[b]{.46\linewidth}
  \centering
  \centerline{\includegraphics[width=\textwidth]{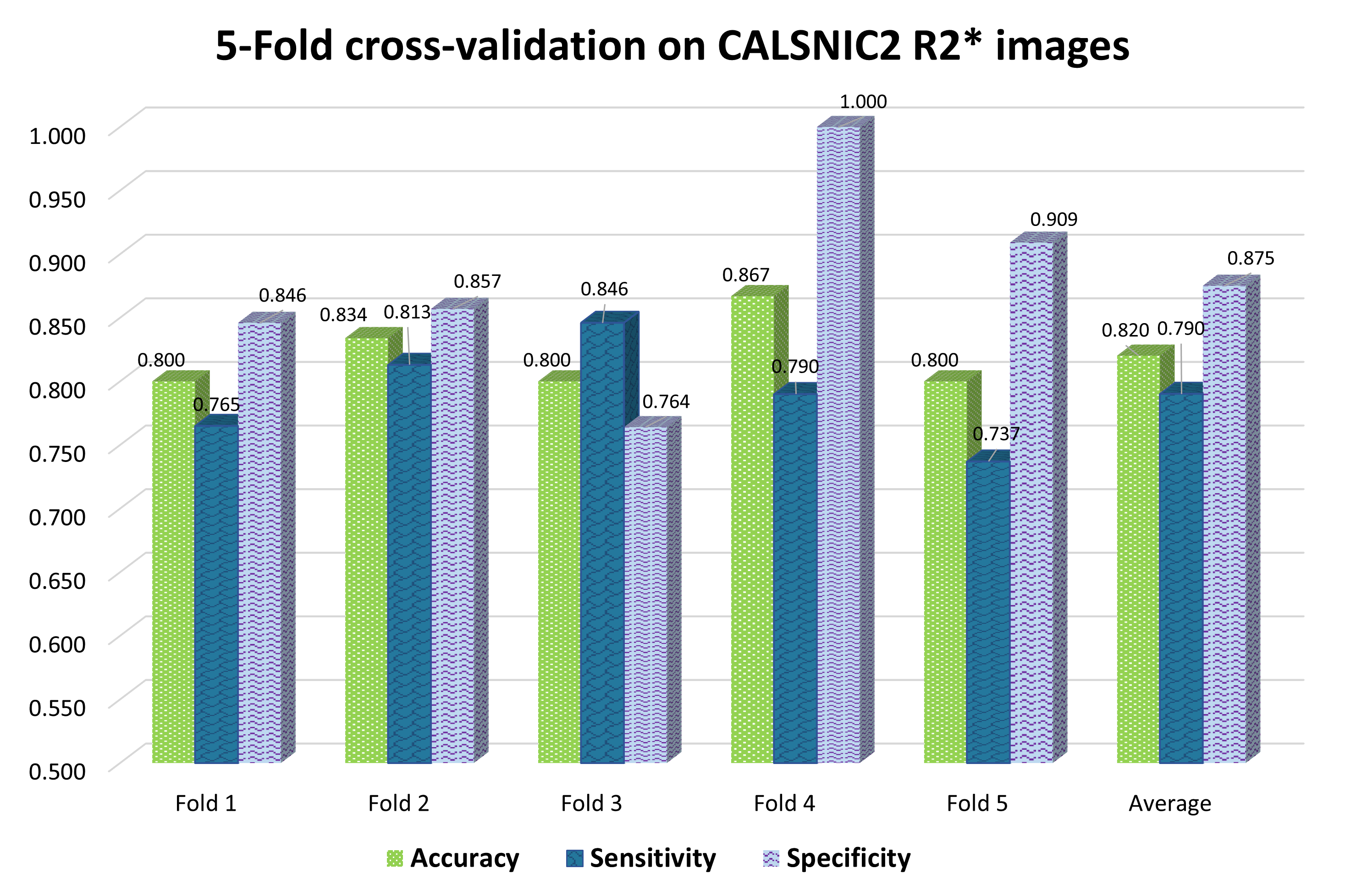}}
  \centerline{(c) CALSNIC2 R2*}\medskip
\end{minipage}
\hfill
\begin{minipage}[b]{0.46\linewidth}
  \centering
  \centerline{\includegraphics[width=\textwidth]{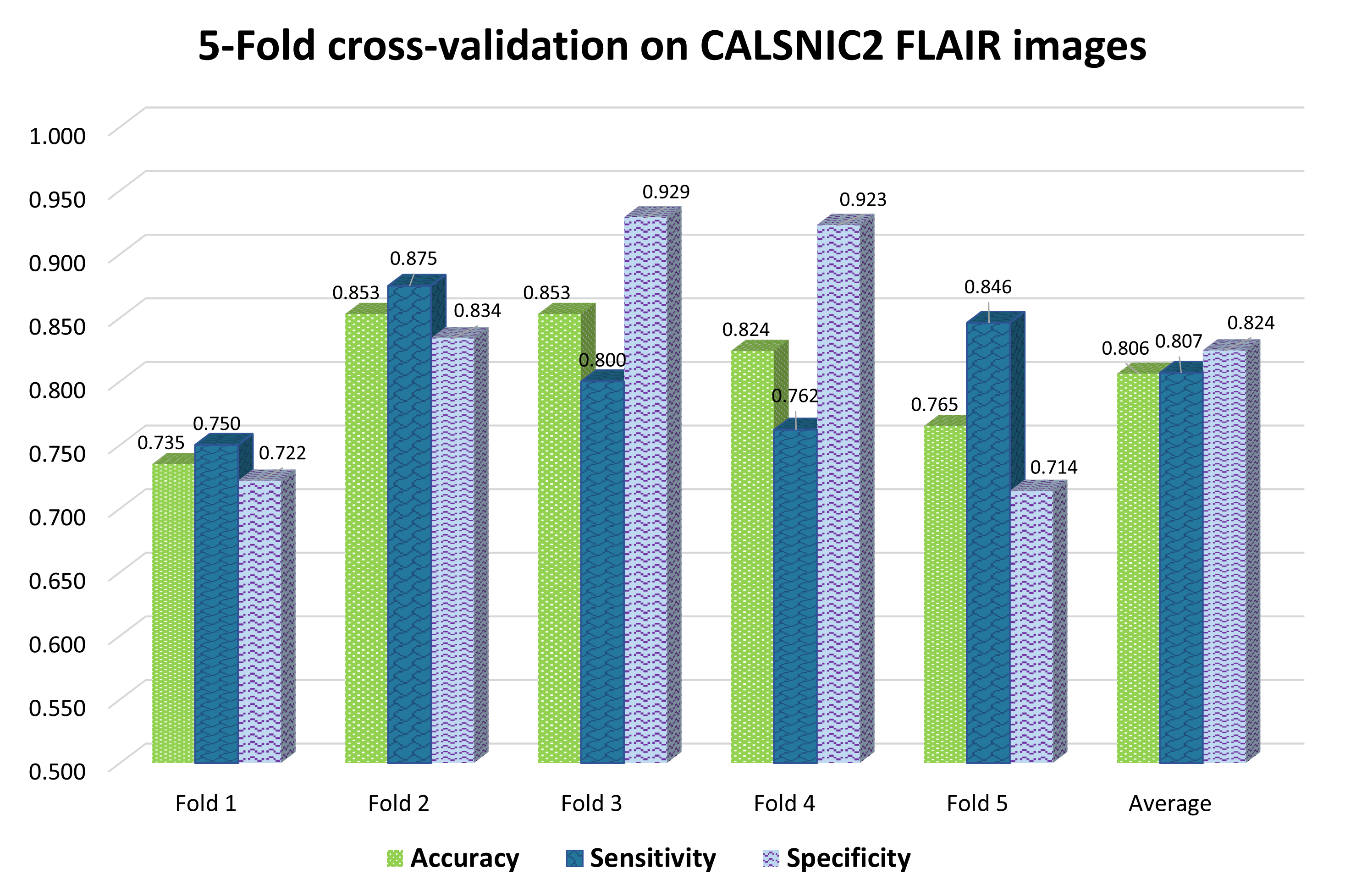}}
  \centerline{(d) CALSNIC2 FLAIR }\medskip
\end{minipage}
\caption{Showing the classification results of the proposed method on different folds for each of the imaging modalities.}
\label{fig:resutls}
\end{figure*}

Commonly used statistical metrics, such as accuracy, sensitivity, specificity, precision, and F1-score are used to assess the classification performance of our proposed method. They are defined in terms of four values which are True Positive (TP), True Negative (TN), False Positive (FP), and False Negative (FN). Sensitivity (SEN), also known as recall, is the capability of a test to identify patients with a disease correctly and is defined as $SEN = \frac{TP}{TP + FN}$. Specificity (SPE), also known as true negative rate, is the ability to determine people without the disorder and is expressed as $SPE= \frac{TN}{TN + FP}$. The positive predictive value or precision (PRE) is the number of relevant items, and high precision implies that an algorithm yields substantially more relevant outcomes than irrelevant ones ($PRE = \frac{TP}{TP + FP}$). The accuracy (ACC) is the fraction of the total number of precisely identified subjects and the total number of samples  in a particular database, which is defined as $ACC = \frac{TP + TN}{TP + TN + FP + FN}$. The F1-score is the harmonic mean of precision and recall which is defined as $F1score = 2 \times \frac{precision \times recall}{precision + recall}$.
Figure \ref{fig:resutls} depicts the outcome of different folds for CALSNIC1 T1-weighted, CALSNIC2 T1-weighted, R2* and FLAIR modality images in terms of ACC, SEN, and SPE as well as displays the average of five folds in the last column of their corresponding chart.

\subsection{Ablation study}
\label{ssec:ablation}
To show the effectiveness of different components and techniques used in our proposed method, we provide an ablation study which is summarized in Table \ref{tab:ablation}. The value of different hyperparameters and the choice of slice selection remain the same for this experiment. We remove one particular function from our proposed framework in the first five settings. First, we show the response from the model without normalizing the pixel intensities of MRI scans. Normalization of the images noticeably boosts the overall accuracy. Then, the importance of data augmentation is shown from the results, where we can see that the accuracy reduced considerably without data augmentation.
Importantly, slice selection plays a vital role in our proposed technique. In particular, including most of the slices of a 3D brain MRI lowers the network's performance remarkably. In the experiment of this w/o slice selection, we choose 120 coronal slices out of 218 (from slice number 51 to 170) as outside of this range does not contain brain and thus any meaningful information. Another influential function of the proposed procedure is leveraging the transfer learning property from the enormous computer vision dataset ImageNet \cite{deng2009imagenet}. Instead of using the pre-trained weights for the ViT and the GFNet networks, training from scratch decreases the accuracy slightly.
Moreover, the idea of applying majority voting at the end of the framework helps to enhance the performance significantly. In other words, bypassing this operation reduces the performance drastically. Finally, we evaluate the performance of our proposed framework's two major building blocks: the ViT and the GFNet. When we apply them independently, they can correctly determine samples which are not identical. By fusing these two networks, we can obtain much-improved classification accuracy in our proposed architecture. In terms of SEN, some other methods outperform our proposed technique while sacrificing the score in SPE and PRE. However, our methodology maintains a balance among all the metrics, which is also reflected in the F1-score.

\begin{table}[htb]
    \scriptsize
	\centering
	\caption{Ablation study for ALS patients vs healthy controls classification on a particular fold of experimentation for the T1-weighted MR images of CALSNIC1 dataset}
	\label{tab:ablation}
	\begin{tabular}{c|ccccc}
		
		\hline
		{\textbf {Method}}  & \textbf {ACC} & \textbf {SEN} & \textbf {SPE} & \textbf {PRE} & \textbf {F1score} \\
		\hline

		\hline
 		\textbf{w/o normalization}  & 0.800 & 0.750 & 0.889 & 0.923 & 0.828   \\

 		\textbf{w/o augmentation}  & 0.760 & \textbf{1.000} & 0.667 & 0.59 & 0.700   \\
        
         \textbf{w/o slice selection}  & 0.680 & 0.647 & 0.750 & 0.846 & 0.734    \\

		\textbf{w/o transfer learning} & 0.840 & 0.846 & 0.834 & 0.846 & 0.846   \\
		
 		\textbf{w/o majority voting} & 0.720 & 0.759 & 0.686 & 0.676 & 0.715   \\
		
		\textbf{ViT only \cite{dosovitskiy2020image}} & 0.800 & 0.834 & 0.769 & 0.769 & 0.800 \\
		
		\textbf{GFNet only \cite{rao2021global}} & 0.840 & 0.800 & 0.917 & 0.923 & 0.857 \\

		\textbf{Proposed method}  & \textbf{0.880} & 0.813 & \textbf{1.000} & \textbf{1.000} & \textbf{0.900} \\
		\hline
		
	\end{tabular}
\end{table}

\subsection{Effects of multi-center study}
\label{ssec:multi-center}
This section  shows the effect of multi-center data or  data acquired with multiple scanners. Especially for deep learning-based models, it becomes much more challenging to learn when the MRI data originates from different scanners \cite{yan2020mri}. Here, we address the classification results from three different setups for the CALSNIC2 T1-weighted images. We take samples from the two largest recruiting centers, namely Toronto and Edmonton, where they use Siemens 3T Prisma model scanner. In the final set, we randomly collect samples from six centers equal to the largest center's dimension. Firstly, we estimate the classification accuracy from the data derived from only the Toronto center, comprising 15 healthy controls and 20 ALS patients. Secondly, we test the classification accuracy of the samples generated from the Edmonton center, which includes 46 normal controls and 35 ALS subjects. Finally, we  randomly assemble MR images from six centers (i.e., Calgary, Edmonton, Toronto, Quebec, Miami, and Montreal) similar to that of the Edmonton center. Table \ref{tab:multicenter} illustrates the assessed score of classification where we can see that when the data comes from a single-center or a single type of scanner with the same image acquisition protocol, the accuracy is higher.
On the other hand, if the data comes from multiple centers or scanners, the performance declines. Another observation is the significance of more training data for deep models. Because of including all the samples from the CALSNIC2 dataset, the classification accuracy improves from approximately 77\% to 82\%.

\begin{table}[htb]
    \scriptsize
	\centering
	\caption{Showing the effects of multi-center study tested on the CALSNIC2 T1-weighted MR images}
	\label{tab:multicenter}
	\begin{tabular}{cc|ccccc}
		\hline
        \multirow{2}{*}{\textbf {Center}}  & \multirow{2}{*}{\textbf {Samples}}  &  \multicolumn{5}{|c}{\textbf {CALSNIC2}}\\
        & & \textbf {ACC} &  \textbf {SEN}    & \textbf{SPE} & \textbf {PRE} & \textbf {F1-score} \\
		\hline

        Toronto  & 35 & 0.813 & 0.900 & 0.708 & 0.800 & 0.845 \\
        
        Edmonton  & 81 & 0.824 & 0.770 & 0.571 & 0.727 & 0.824\\
        
        Multi-center & 81 & 0.765 & 1.000 & 0.714 & 0.600 & 0.765\\

		\hline
	\end{tabular}
\end{table}

\subsection{Effects of different MRI modalities}
\label{ssec:multi-modal}
Our proposed framework investigates the potential of some neuroimaging modalities in the application of ALS disease classification. Firstly, we consider T1-weighted images of the CALSNIC1 and CALSNIC2 datasets as T1-weighted scans are commonly used for neurodegenerative disorder classification based on structural metrics such as volumes. Secondly, we evaluate the performance of the R2* modality of the CALSNIC2 dataset, which is rarely explored for ALS classification. Finally, the classification accuracy is calculated with respect to the FLAIR imaging modality on the CALSNIC2 dataset. Table \ref{tab:multimodal} reflects the performance of our multi-modal imaging analysis evaluation. Out of the five evaluation metrics, this study reveals that the R2* modality achieves better results in terms of accuracy, specificity, precision and F1-score. The classification accuracy of T1-weighted images is very close to the performance of R2* as well as obtains the highest sensitivity. However, the FLAIR modality attains slightly lower classification accuracy compared to other modalities.
\begin{table*}[htb]
    \footnotesize
	\centering
	\caption{Showing the classification results of multi-modal MRI study}
	\label{tab:multimodal}
	\begin{tabular}{cc|ccccc}
		\hline
        {\textbf {Modality(Dataset)}}   & {\textbf {Samples}}
        & \textbf {ACC} &  \textbf {SEN} & \textbf{SPE} & \textbf{PRE} & \textbf{F1-score} \\
		\hline

        T1-W (CALSNIC1) & 120 & 0.816 & \textbf{0.843} & 0.812 & 0.800 & 0.815 \\

        T1-W (CALSNIC2) & 223 & 0.818 & 0.824 & 0.815 & 0.800 & 0.811 \\

        R2* (CALSNIC2) & 148 & \textbf{0.820} & 0.790 & \textbf{0.875} & \textbf{0.880} & \textbf{0.829} \\

        FLAIR (CALSNIC2) & 168 & 0.806 & 0.807 & 0.824 & 0.812 & 0.803\\
        
		\hline
	\end{tabular}
\end{table*}

\subsection{Effects of slice selection}
\label{ssec:slice_selection}
\begin{figure*}[htb]

\begin{minipage}[b]{.46\linewidth}
  \centering
  \centerline{\includegraphics[width=\textwidth]{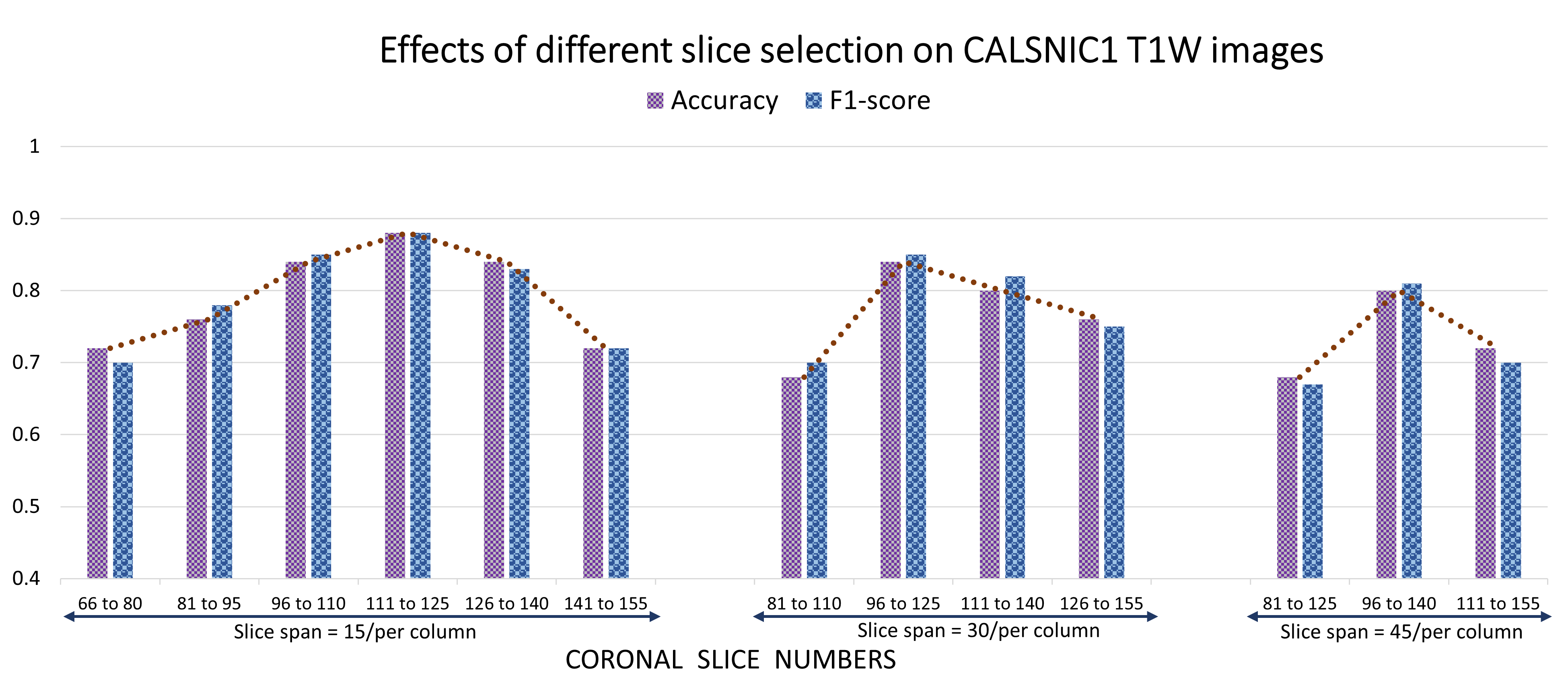}}
  \centerline{(a) CALSNIC1 T1W}\medskip
\end{minipage}
\hfill
\begin{minipage}[b]{0.46\linewidth}
  \centering
  \centerline{\includegraphics[width=\textwidth]{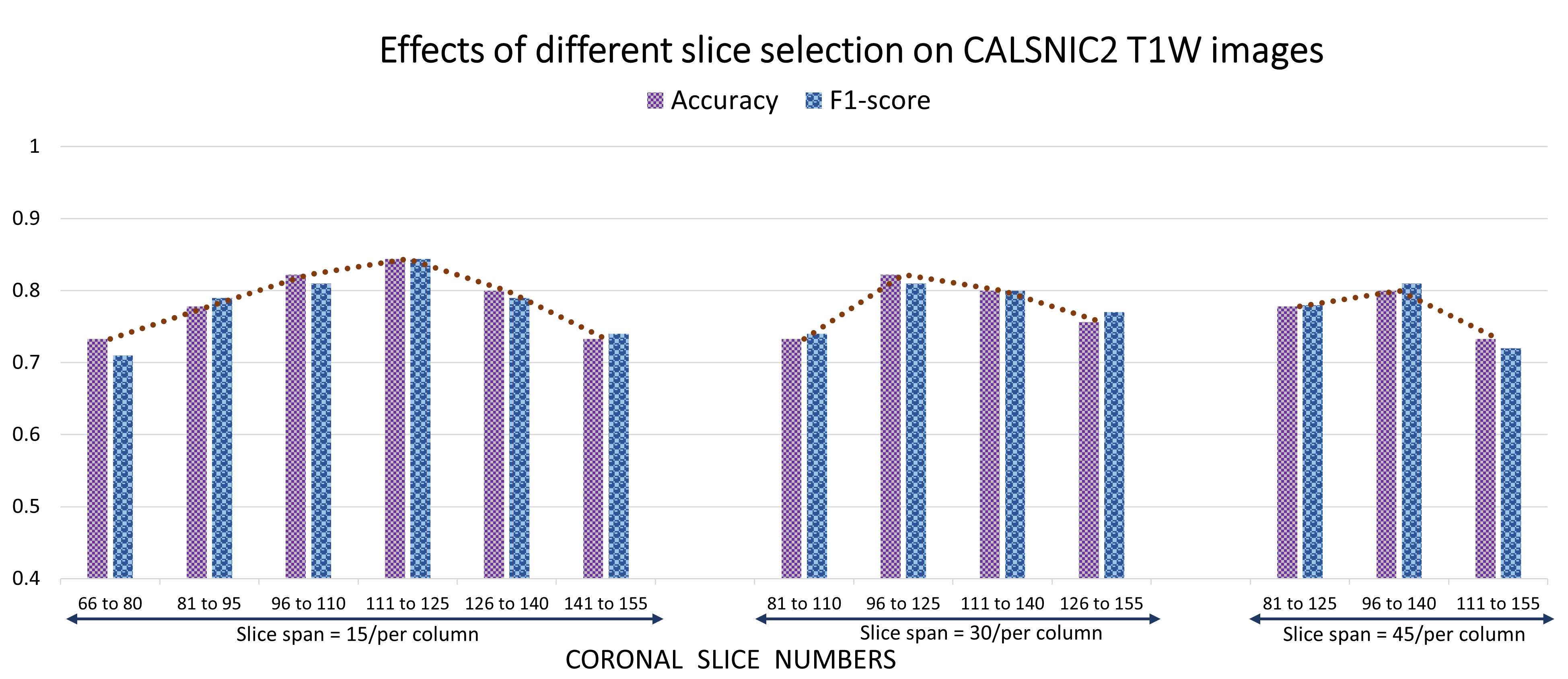}}
  \centerline{(b) CALSNIC2 T1W}\medskip
\end{minipage}

\begin{minipage}[b]{.46\linewidth}
  \centering
  \centerline{\includegraphics[width=\textwidth]{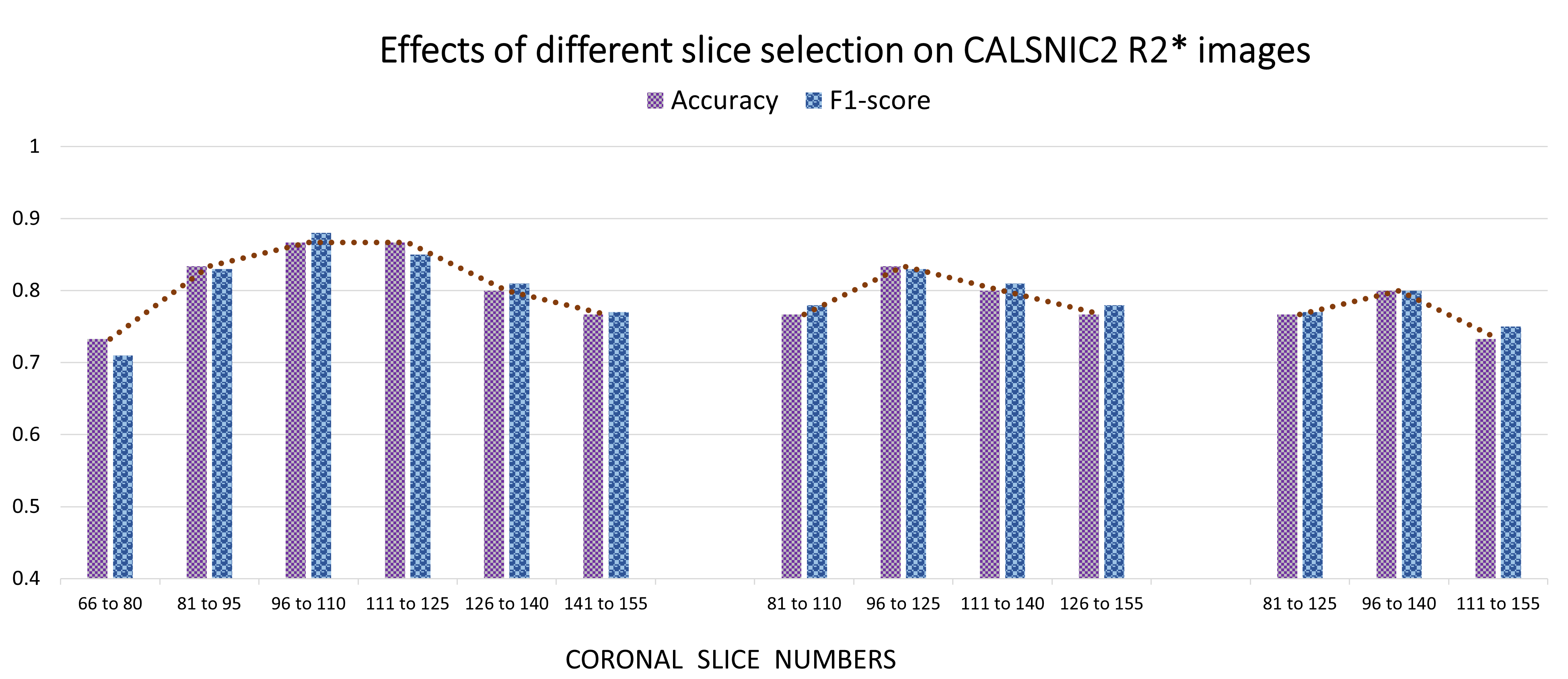}}
  \centerline{(c) CALSNIC2 R2*}\medskip
\end{minipage}
\hfill
\begin{minipage}[b]{0.46\linewidth}
  \centering
  \centerline{\includegraphics[width=\textwidth]{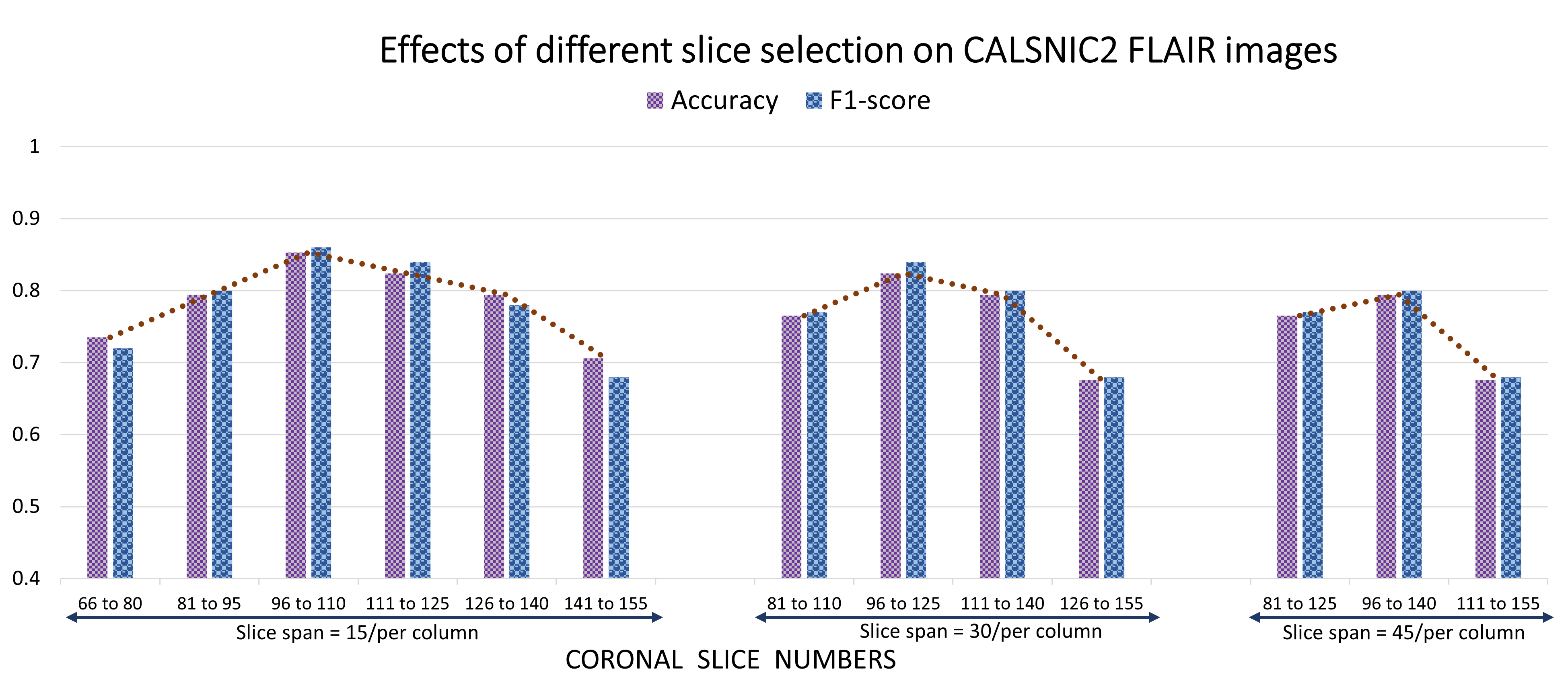}}
  \centerline{(d) CALSNIC2 FLAIR }\medskip
\end{minipage}
\caption{Showing the classification result effects on different range of coronal slice selection for each of the MRI modalities used in our study.}
\label{fig:slice_selection}
\end{figure*}

\begin{figure}[h!]

\begin{minipage}[b]{1.0\linewidth}
  \centering
  \centerline{\includegraphics[width=\textwidth]{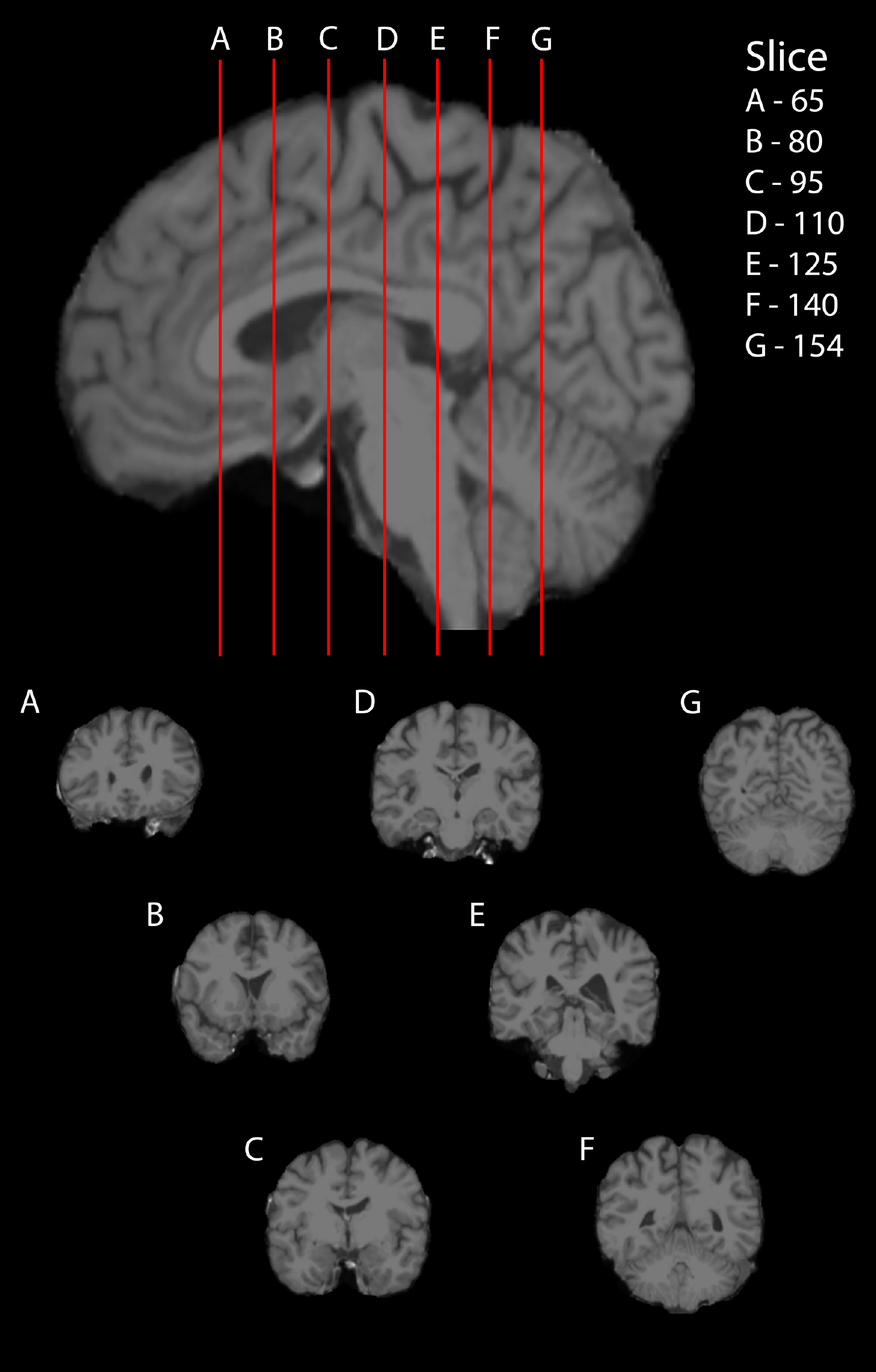}}
\end{minipage}
\caption{Slice number and the corresponding coronal MR image from a T1-weighted scan from an ALS patient. Out of 218 coronal slices, the best performance is found from the slice range of 111 (D) to 125 (E) for the T1-weighted images.}
\label{fig:slice_anatomy}
\end{figure}

After applying the FreeSurfer $autorecon1$, FSL $flirt$, and resize commands respectively the reconstructed image size becomes $224 \times 218 \times 224$. So, each coronal slice (total 218) has a dimension of $224 \times 224$. From the 2D slice view perspective, there is no meaningful information available at the beginning or the ending part of the volume. In other words, most of the tissues or important structural information can be found in the central part of the volume. For that reason, we explore the effectiveness of a wide range of slices to investigate which part of the volume provides the best performance, and the results are demonstrated in Fig \ref{fig:slice_selection}. We start with an interval of 15 consecutive slices from the central slice location and analyze 45 slices in the forward and backward directions. Then, we experiment with different combinations of successive slices within these 90 slice spans. For example, we are combining 30 or 45 consecutive slices. After careful observation, the best performance is found from the slice range of 111 to 125 for the T1-weighted images. The closest result to the best performance comes from the slice span of 96 to 110. Increasing the number of slices for training could not boost the performance, as noticed in each of the chart's middle and right segments of Fig \ref{fig:slice_selection}. Moreover, increasing the number of slices for model training also noticeably accelerates the total training time. However, the best classification accuracy yielded from the R2* maps and FLAIR images is from the slice range of 96 to 110. The slices from 111 to 125 also accomplish a comparable job in classifying ALS patients with R2* and FLAIR images. The anatomical features expected in different slice ranges can be perceived in Fig. \ref{fig:slice_anatomy}.

\subsection{Testing on an unknown data center}
\label{ssec:unknown_center}
This section exemplifies our proposed method's generalization ability and robustness by testing on an unknown database. The Utah center in CALSNIC2 comprises nine healthy controls and ALS patients. So far, we have not incorporated this site's MRI data into our model's training processes. Our trained model of CALSNIC2 T1-weighted images scored seven correct predictions out of these nine subjects, one misclassification from the control group and the other from the patient class. As a result, an accuracy of 0.778, sensitivity 0.857 and F1-score 0.857 are achieved through our proposed network in classifying these two groups. 

\subsection{Comparison}
\label{ssec:comparison}
We compare our proposed network with some popular deep learning-based 2D and 3D architectures and highlight the results in table \ref{tab:compare}. Moreover, we also reproduce the classification accuracy of one previous state-of-the-art work, M-CoHOG \cite{elahi2020texture} in ALS classification for the CALSNIC1 dataset. As it requires a manual slice selection process from the experts, we could not report their accuracy for the CALSNIC2 database. Firstly, we estimate the classification accuracy for the widely used ResNet architecture \cite{he2016deep} with a depth of different layers such as 10, 18, 50, 101, 152 and report the best accuracy among them. Secondly, we evaluate the performance of the MobileNet network \cite{howard2017mobilenets} that uses depthwise separable convolutions to develop lightweight deep neural networks. Thirdly, we measure the accuracy with ShuffleNet framework \cite{zhang2018shufflenet}, which utilizes pointwise group convolution and channel shuffle to provide efficient computation cost. Next, we calculate the performance with another popular deep model named EfficientNet \cite{tan2019efficientnet}, which can balance the network's depth, width, and resolution effectively. For the 2D CNN-based architectures, we follow similar steps and data, such as input slices and majority voting, as in our proposed method. The input for the 3D-based framework is the whole 3D brain MRI to carry out the classification outcome. The average classification accuracies and different hyper-parameters values selected for the baseline methods are also attained from the Stratified five-fold CV, as mentioned earlier. In table \ref{tab:compare}, our proposed strategy outperformed all other approaches in most of the evaluation metrics except SPE. MobileNet architecture achieves higher SPE while producing inferior SEN, which indicates the model is biased towards a particular class.

\begin{table*}[htb]
    \footnotesize
	\centering
	\caption{Comparison with some popular CNN architectures and previous work for ALS patients vs. healthy controls classification tested on the T1-weighted MR images of CALSNIC1 and CALSNIC2 datasets.}
	\label{tab:compare}
	\begin{tabular}{c|cccc|cccc}
		
		\hline
        \multirow{2}{*}{\textbf {Reference}}  & \multicolumn{4}{c}{\textbf {CALSNIC1}} & \multicolumn{4}{|c}{\textbf {CALSNIC2}} \\
        & \textbf {ACC} &  \textbf {SEN}    & \textbf{SPE}  & \textbf{F1-score} & \textbf {ACC} &  \textbf {SEN}  & \textbf{SPE}  & \textbf{F1-score} \\
		\hline

        ResNet (2D) \cite{he2016deep}  & 0.717 & 0.631 & 0.811 & 0.699 & 0.726 & 0.667 & 0.783 & 0.746\\
        
        ResNet (3D) \cite{he2016deep}  & 0.720 & 0.692 & 0.750 & 0.720 & 0.733 & 0.727 & 0.739 & 0.733\\
        
        ShuffleNet (2D) \cite{zhang2018shufflenet} &  0.680  & 0.580 & 0.789 & 0.654 & 0.696 & 0.697 & 0.696 & 0.692\\
        
        ShuffleNet (3D) \cite{zhang2018shufflenet} &  0.680  & 0.692 & 0.667 & 0.692 & 0.689 & 0.773 & 0.609 & 0.708\\
        
        MobileNet (2D) \cite{howard2017mobilenets}  &  0.667 & 0.441 & 0.911 & 0.580 & 0.704 & 0.546 & 0.783 & 0.643\\
        
        MobileNet (3D) \cite{howard2017mobilenets}  & 0720 & 0.539 & \textbf {0.917} & 0.667 & 0.711 & 0.546 & \textbf {0.870} & 0.649\\
        
        EfficientNet (3D) \cite{tan2019efficientnet}  &  0.680  & 0.667 & 0.692 & 0.667 & 0.711 & 0.727 & 0.696 & 0.696\\
        
        M-CoHOG (2D) \cite{elahi2020texture}   & 0.745 & 0.786 & 0.688  & 0.752 & - & -  & - & - \\

		{\textbf {Proposed method}}  & {\textbf {0.816}} & {\textbf {0.843}} & 0.812  & {\textbf {0.815}} & {\textbf {0.818}} & {\textbf {0.824}} & 0.815  & {\textbf {0.811}}\\

		\hline
		
	\end{tabular}
\end{table*}

\section{Discussion}
\label{sec:discussion}
The primary challenge to classify ALS subjects from normal controls using a deep model is the availability of adequate neuroimaging data to train the model. Secondly, the variability of data because of different MRI scanners in distinct locations and scanning protocols also creates challenges in analyzing and classifying the resulting images. In this latter case, the distribution of imaging data becomes slightly dissimilar and results in inconsistent performance in deep networks, usually referred to as the multi-center problem. Finally, the etiology and phenotype of the disease are so complex and diverse that there is no transparent gold-standard region of interest that can always reflect the distinguishing pathological feature causing ALS.

This study introduces an efficient vision transformer-based deep framework by fusing both spatial and frequency domain features. We evaluate the performance of our proposed model for the classification of healthy control participants from the complex neurodegenerative disorder ALS subjects. After testing different MRI modalities, we have found slightly better performance from R2* images. However, the classification accuracy of T1-weighted and FLAIR scans is also comparable to R2*. The proposed method is robust to multi-center data as well as outperformed many prior CNN-based deep models.

Because of the limited number of samples, we excluded the Utah center from the CALSNIC2 dataset for basic experiments to alleviate the variability of data distribution in our training process. However, we analyze the data of this center in a separate experiment to see how our model performs on an unknown center. Moreover, the FreeSurfer \cite{fischl2012freesurfer} could not perform the skull stripping satisfactorily on some samples for R2* and FLAIR modality, which makes the total sample size slightly smaller in R2* and FLAIR compared to that of CALSNIC2 T1-weighted data. In addition to the manual observation, we confirm the quality of each image using an open-source quality control tool for MRI data named MRQy \cite{sadri2020mrqy}. MRQy depicts different quality-related metrics such as peak signal-to-noise ratio (PSNR), contrast-to-noise ratio (CNR), coefficient of variation of the foreground patch (CVP) for shading artifacts, entropy focus criterion (EFC) for motion artifacts etc. Their graphical interface makes it easier to identify the presence of any outliers or inconsistencies in samples in a dataset.

The preprocessing steps we execute in our proposed workflow are minimal, free from any manual parameter value selection and easy to reproduce. The slight difference in the performance of different choices of slices for CALSNIC1 T1-weighted and CALSNIC2 T1-weighted images may be due to the difference in the number of subjects used to train the model. In contrast, the other cause could be the difference in the image acquisition protocols. More specifically, in CALSNIC1, the acquisition orientation is axial, but in CALSNIC2, the acquisition orientation is sagittal.

Recently, we published another vision transformer-based deep model titled ADDFormer \cite{kushol2022addformer} for classifying Alzheimer's patients from normal controls using structural MRI data. However, there are some notable differences in $SF^2Former$ compared to ADDFormer. Firstly, the fusion part in ADDFormer and $SF^2Former$ is significantly different, whereas the former model uses a third transformer model to merge the features. Secondly, due to having the third transformer block, the number of parameters and computational costs are comparatively higher in ADDFormer. Thirdly, the process of slice selection was different in ADDFormer architecture.

\section{Conclusion and Future Work}
\label{sec:conclusion}
This study thoroughly investigates the potential of ViT architecture integrated with spatial and frequency domain features in a novel manner for a complex neurodegenerative disease classification task that distinguishes ALS patients from healthy controls. The classification accuracy of our proposed network has outperformed prior popular deep models and achieved a satisfactory performance using T1-weighted, FLAIR and R2* MR imaging data. However, the best performance we observe among these modalities is from the R2* maps which suggests further exploration is needed to use it effectively for ALS diagnosis. Our introduced methodology will bring MRI closer to the reality of providing biomarkers for ALS diagnosis and monitoring disease progression as well as response to therapy. We plan to incorporate clinical features and imaging data to enhance classification performance in the future. Other neuroimaging modalities like functional MRI (fMRI) and DTI can also be investigated with a similar framework. Our proposed architecture is flexible to adapt other neurodegenerative disease classification tasks with appropriate slice selection where frequency domain information plays an essential role in feature extraction.

\section*{Declaration of Competing Interest}
\label{sec:competing_interest}
The authors state that they have no known competing financial interests or personal relationships that may have influenced the work documented in this manuscript.

\section*{Ethics statement}
\label{sec:ethics}
The study was conducted in acceptance with the approval of the research ethics boards across all the CALSNIC sites. The patients/participants gave their written informed consent to engage in this study.

\section*{Data and code availability statement}
\label{sec:data_code}
The Department of Medicine at the University of Alberta stores all of the neuroimaging data used in this investigation. All processed data used in the analysis are available to the corresponding author upon reasonable request, following the University of Alberta Ethics Committee's data sharing and privacy rules. The preprocessing of the MRI data has been performed with open-source tools (FSL and FreeSurfer). Our coding deeply follows the publicly available implementation of the ViT(\url{https://github.com/jeonsworld/ViT-pytorch}) and the GFNet(\url{https://github.com/raoyongming/GFNet}). The modified code to run our proposed framework will be made available to a GitHub repository after the acceptance of the paper. The implementation of the 3D CNN methods is available here(\url{https://github.com/xmuyzz/3D-CNN-PyTorch}).

\section*{Acknowledgments}
We would like to thank all the participants in the CALSNIC study and their caregivers. This study has been supported by the ALS Society of Canada, Brain Canada, Natural Sciences and Engineering Research Council of Canada (NSERC), and Prime Minister Fellowship Bangladesh.

\bibliographystyle{model2-names.bst}\biboptions{authoryear}
\bibliography{als, transformer}

\begin{thebibliography}{44}
\expandafter\ifx\csname natexlab\endcsname\relax\def\natexlab#1{#1}\fi
\providecommand{\url}[1]{\texttt{#1}}
\providecommand{\href}[2]{#2}
\providecommand{\path}[1]{#1}
\providecommand{\DOIprefix}{doi:}
\providecommand{\ArXivprefix}{arXiv:}
\providecommand{\URLprefix}{URL: }
\providecommand{\Pubmedprefix}{pmid:}
\providecommand{\doi}[1]{\href{http://dx.doi.org/#1}{\path{#1}}}
\providecommand{\Pubmed}[1]{\href{pmid:#1}{\path{#1}}}
\providecommand{\bibinfo}[2]{#2}
\ifx\xfnm\relax \def\xfnm[#1]{\unskip,\space#1}\fi
\bibitem[{Cai et~al.(2022)Cai, He, Lin and Tang}]{cai2022uni4eye}
\bibinfo{author}{Cai, Z.}, \bibinfo{author}{He, H.}, \bibinfo{author}{Lin, L.},
  \bibinfo{author}{Tang, X.}, \bibinfo{year}{2022}.
\newblock \bibinfo{title}{Uni4eye: Unified 2d and 3d self-supervised
  pre-training via masked image modeling transformer for ophthalmic image
  classification}.
\newblock \bibinfo{journal}{arXiv preprint arXiv:2203.04614} .
\bibitem[{Chen et~al.(2022)Chen, Frey, He, Segars, Li and
  Du}]{chen2022transmorph}
\bibinfo{author}{Chen, J.}, \bibinfo{author}{Frey, E.C.}, \bibinfo{author}{He,
  Y.}, \bibinfo{author}{Segars, W.P.}, \bibinfo{author}{Li, Y.},
  \bibinfo{author}{Du, Y.}, \bibinfo{year}{2022}.
\newblock \bibinfo{title}{Transmorph: Transformer for unsupervised medical
  image registration}.
\newblock \bibinfo{journal}{Medical Image Analysis} \bibinfo{volume}{82},
  \bibinfo{pages}{102615}.
\bibitem[{Chen et~al.(2021)Chen, Lu, Yu, Luo, Adeli, Wang, Lu, Yuille and
  Zhou}]{chen2021transunet}
\bibinfo{author}{Chen, J.}, \bibinfo{author}{Lu, Y.}, \bibinfo{author}{Yu, Q.},
  \bibinfo{author}{Luo, X.}, \bibinfo{author}{Adeli, E.},
  \bibinfo{author}{Wang, Y.}, \bibinfo{author}{Lu, L.},
  \bibinfo{author}{Yuille, A.L.}, \bibinfo{author}{Zhou, Y.},
  \bibinfo{year}{2021}.
\newblock \bibinfo{title}{Transunet: Transformers make strong encoders for
  medical image segmentation}.
\newblock \bibinfo{journal}{arXiv preprint arXiv:2102.04306} .
\bibitem[{Chen et~al.(2020)Chen, Zhang, Huang and
  Chen}]{chen2020identification}
\bibinfo{author}{Chen, Q.F.}, \bibinfo{author}{Zhang, X.H.},
  \bibinfo{author}{Huang, N.X.}, \bibinfo{author}{Chen, H.J.},
  \bibinfo{year}{2020}.
\newblock \bibinfo{title}{Identification of amyotrophic lateral sclerosis based
  on diffusion tensor imaging and support vector machine}.
\newblock \bibinfo{journal}{Frontiers in neurology} \bibinfo{volume}{11},
  \bibinfo{pages}{275}.
\bibitem[{Cheng et~al.(2020)Cheng, Dalca, Fischl, Z{\"o}llei, Initiative
  et~al.}]{cheng2020cortical}
\bibinfo{author}{Cheng, J.}, \bibinfo{author}{Dalca, A.V.},
  \bibinfo{author}{Fischl, B.}, \bibinfo{author}{Z{\"o}llei, L.},
  \bibinfo{author}{Initiative, A.D.N.}, et~al., \bibinfo{year}{2020}.
\newblock \bibinfo{title}{Cortical surface registration using unsupervised
  learning}.
\newblock \bibinfo{journal}{NeuroImage} \bibinfo{volume}{221},
  \bibinfo{pages}{117161}.
\bibitem[{Deng et~al.(2009)Deng, Dong, Socher, Li, Li and
  Fei-Fei}]{deng2009imagenet}
\bibinfo{author}{Deng, J.}, \bibinfo{author}{Dong, W.},
  \bibinfo{author}{Socher, R.}, \bibinfo{author}{Li, L.J.},
  \bibinfo{author}{Li, K.}, \bibinfo{author}{Fei-Fei, L.},
  \bibinfo{year}{2009}.
\newblock \bibinfo{title}{Imagenet: A large-scale hierarchical image database},
  in: \bibinfo{booktitle}{2009 IEEE conference on computer vision and pattern
  recognition}, \bibinfo{organization}{Ieee}. pp. \bibinfo{pages}{248--255}.
\bibitem[{Dosovitskiy et~al.(2020)Dosovitskiy, Beyer, Kolesnikov, Weissenborn,
  Zhai, Unterthiner, Dehghani, Minderer, Heigold, Gelly
  et~al.}]{dosovitskiy2020image}
\bibinfo{author}{Dosovitskiy, A.}, \bibinfo{author}{Beyer, L.},
  \bibinfo{author}{Kolesnikov, A.}, \bibinfo{author}{Weissenborn, D.},
  \bibinfo{author}{Zhai, X.}, \bibinfo{author}{Unterthiner, T.},
  \bibinfo{author}{Dehghani, M.}, \bibinfo{author}{Minderer, M.},
  \bibinfo{author}{Heigold, G.}, \bibinfo{author}{Gelly, S.}, et~al.,
  \bibinfo{year}{2020}.
\newblock \bibinfo{title}{An image is worth 16x16 words: Transformers for image
  recognition at scale}.
\newblock \bibinfo{journal}{arXiv preprint arXiv:2010.11929} .
\bibitem[{Elahi et~al.(2020)Elahi, Kalra, Zinman, Genge, Korngut and
  Yang}]{elahi2020texture}
\bibinfo{author}{Elahi, G.M.E.}, \bibinfo{author}{Kalra, S.},
  \bibinfo{author}{Zinman, L.}, \bibinfo{author}{Genge, A.},
  \bibinfo{author}{Korngut, L.}, \bibinfo{author}{Yang, Y.H.},
  \bibinfo{year}{2020}.
\newblock \bibinfo{title}{Texture classification of mr images of the brain in
  als using m-cohog: A multi-center study}.
\newblock \bibinfo{journal}{Computerized Medical Imaging and Graphics}
  \bibinfo{volume}{79}, \bibinfo{pages}{101659}.
\bibitem[{Fabes et~al.(2017)Fabes, Matthews, Filippini, Talbot, Jenkinson and
  Turner}]{fabes2017quantitative}
\bibinfo{author}{Fabes, J.}, \bibinfo{author}{Matthews, L.},
  \bibinfo{author}{Filippini, N.}, \bibinfo{author}{Talbot, K.},
  \bibinfo{author}{Jenkinson, M.}, \bibinfo{author}{Turner, M.R.},
  \bibinfo{year}{2017}.
\newblock \bibinfo{title}{Quantitative flair mri in amyotrophic lateral
  sclerosis}.
\newblock \bibinfo{journal}{Academic radiology} \bibinfo{volume}{24},
  \bibinfo{pages}{1187--1194}.
\bibitem[{Fischl(2012)}]{fischl2012freesurfer}
\bibinfo{author}{Fischl, B.}, \bibinfo{year}{2012}.
\newblock \bibinfo{title}{Freesurfer}.
\newblock \bibinfo{journal}{Neuroimage} \bibinfo{volume}{62},
  \bibinfo{pages}{774--781}.
\bibitem[{He et~al.(2016)He, Zhang, Ren and Sun}]{he2016deep}
\bibinfo{author}{He, K.}, \bibinfo{author}{Zhang, X.}, \bibinfo{author}{Ren,
  S.}, \bibinfo{author}{Sun, J.}, \bibinfo{year}{2016}.
\newblock \bibinfo{title}{Deep residual learning for image recognition}, in:
  \bibinfo{booktitle}{Proceedings of the IEEE conference on computer vision and
  pattern recognition}, pp. \bibinfo{pages}{770--778}.
\bibitem[{Hecht et~al.(2001)Hecht, Fellner, Fellner, Hilz, Heuss and
  Neund{\"o}rfer}]{hecht2001mri}
\bibinfo{author}{Hecht, M.}, \bibinfo{author}{Fellner, F.},
  \bibinfo{author}{Fellner, C.}, \bibinfo{author}{Hilz, M.},
  \bibinfo{author}{Heuss, D.}, \bibinfo{author}{Neund{\"o}rfer, B.},
  \bibinfo{year}{2001}.
\newblock \bibinfo{title}{Mri-flair images of the head show corticospinal tract
  alterations in als patients more frequently than t2-, t1-and
  proton-density-weighted images}.
\newblock \bibinfo{journal}{Journal of the neurological sciences}
  \bibinfo{volume}{186}, \bibinfo{pages}{37--44}.
\bibitem[{Hendrycks and Gimpel(2016)}]{hendrycks2016gaussian}
\bibinfo{author}{Hendrycks, D.}, \bibinfo{author}{Gimpel, K.},
  \bibinfo{year}{2016}.
\newblock \bibinfo{title}{Gaussian error linear units (gelus)}.
\newblock \bibinfo{journal}{arXiv preprint arXiv:1606.08415} .
\bibitem[{Howard et~al.(2017)Howard, Zhu, Chen, Kalenichenko, Wang, Weyand,
  Andreetto and Adam}]{howard2017mobilenets}
\bibinfo{author}{Howard, A.G.}, \bibinfo{author}{Zhu, M.},
  \bibinfo{author}{Chen, B.}, \bibinfo{author}{Kalenichenko, D.},
  \bibinfo{author}{Wang, W.}, \bibinfo{author}{Weyand, T.},
  \bibinfo{author}{Andreetto, M.}, \bibinfo{author}{Adam, H.},
  \bibinfo{year}{2017}.
\newblock \bibinfo{title}{Mobilenets: Efficient convolutional neural networks
  for mobile vision applications}.
\newblock \bibinfo{journal}{arXiv preprint arXiv:1704.04861} .
\bibitem[{Ignjatovi{\'c} et~al.(2013)Ignjatovi{\'c}, Stevi{\'c}, Lavrni{\'c},
  Dakovi{\'c} and Ba{\v{c}}i{\'c}}]{ignjatovic2013brain}
\bibinfo{author}{Ignjatovi{\'c}, A.}, \bibinfo{author}{Stevi{\'c}, Z.},
  \bibinfo{author}{Lavrni{\'c}, S.}, \bibinfo{author}{Dakovi{\'c}, M.},
  \bibinfo{author}{Ba{\v{c}}i{\'c}, G.}, \bibinfo{year}{2013}.
\newblock \bibinfo{title}{Brain iron mri: a biomarker for amyotrophic lateral
  sclerosis}.
\newblock \bibinfo{journal}{Journal of magnetic resonance imaging}
  \bibinfo{volume}{38}, \bibinfo{pages}{1472--1479}.
\bibitem[{Jaiswal(2019)}]{jaiswal2019riluzole}
\bibinfo{author}{Jaiswal, M.K.}, \bibinfo{year}{2019}.
\newblock \bibinfo{title}{Riluzole and edaravone: A tale of two amyotrophic
  lateral sclerosis drugs}.
\newblock \bibinfo{journal}{Medicinal Research Reviews} \bibinfo{volume}{39},
  \bibinfo{pages}{733--748}.
\bibitem[{Jenkinson et~al.(2012)Jenkinson, Beckmann, Behrens, Woolrich and
  Smith}]{jenkinson2012fsl}
\bibinfo{author}{Jenkinson, M.}, \bibinfo{author}{Beckmann, C.F.},
  \bibinfo{author}{Behrens, T.E.}, \bibinfo{author}{Woolrich, M.W.},
  \bibinfo{author}{Smith, S.M.}, \bibinfo{year}{2012}.
\newblock \bibinfo{title}{Fsl}.
\newblock \bibinfo{journal}{Neuroimage} \bibinfo{volume}{62},
  \bibinfo{pages}{782--790}.
\bibitem[{Jin et~al.(2016)Jin, Hu, Zhang, Jia and Dang}]{jin2016hyperintensity}
\bibinfo{author}{Jin, J.}, \bibinfo{author}{Hu, F.}, \bibinfo{author}{Zhang,
  Q.}, \bibinfo{author}{Jia, R.}, \bibinfo{author}{Dang, J.},
  \bibinfo{year}{2016}.
\newblock \bibinfo{title}{Hyperintensity of the corticospinal tract on flair: A
  simple and sensitive objective upper motor neuron degeneration marker in
  clinically verified amyotrophic lateral sclerosis}.
\newblock \bibinfo{journal}{Journal of the neurological sciences}
  \bibinfo{volume}{367}, \bibinfo{pages}{177--183}.
\bibitem[{Kalra et~al.(2020)Kalra, Khan, Barlow, Beaulieu, Benatar, Briemberg,
  Chenji, Clua, Das, Dionne et~al.}]{kalra2020canadian}
\bibinfo{author}{Kalra, S.}, \bibinfo{author}{Khan, M.},
  \bibinfo{author}{Barlow, L.}, \bibinfo{author}{Beaulieu, C.},
  \bibinfo{author}{Benatar, M.}, \bibinfo{author}{Briemberg, H.},
  \bibinfo{author}{Chenji, S.}, \bibinfo{author}{Clua, M.G.},
  \bibinfo{author}{Das, S.}, \bibinfo{author}{Dionne, A.}, et~al.,
  \bibinfo{year}{2020}.
\newblock \bibinfo{title}{The canadian als neuroimaging consortium (calsnic)-a
  multicentre platform for standardized imaging and clinical studies in als}.
\newblock \bibinfo{journal}{MedRxiv} .
\bibitem[{Kocar et~al.(2021)Kocar, Behler, Ludolph, M{\"u}ller and
  Kassubek}]{kocar2021multiparametric}
\bibinfo{author}{Kocar, T.D.}, \bibinfo{author}{Behler, A.},
  \bibinfo{author}{Ludolph, A.C.}, \bibinfo{author}{M{\"u}ller, H.P.},
  \bibinfo{author}{Kassubek, J.}, \bibinfo{year}{2021}.
\newblock \bibinfo{title}{Multiparametric microstructural mri and machine
  learning classification yields high diagnostic accuracy in amyotrophic
  lateral sclerosis: proof of concept}.
\newblock \bibinfo{journal}{Frontiers in neurology} \bibinfo{volume}{12}.
\bibitem[{Kushol et~al.(2022)Kushol, Masoumzadeh, Huo, Kalra and
  Yang}]{kushol2022addformer}
\bibinfo{author}{Kushol, R.}, \bibinfo{author}{Masoumzadeh, A.},
  \bibinfo{author}{Huo, D.}, \bibinfo{author}{Kalra, S.},
  \bibinfo{author}{Yang, Y.H.}, \bibinfo{year}{2022}.
\newblock \bibinfo{title}{Addformer: Alzheimer’s disease detection from
  structural mri using fusion transformer}, in: \bibinfo{booktitle}{2022 IEEE
  19th International Symposium on Biomedical Imaging (ISBI)},
  \bibinfo{organization}{IEEE}. pp. \bibinfo{pages}{1--5}.
\bibitem[{Kushol and Salekin(2020)}]{kushol2020rbvs}
\bibinfo{author}{Kushol, R.}, \bibinfo{author}{Salekin, M.S.},
  \bibinfo{year}{2020}.
\newblock \bibinfo{title}{Rbvs-net: A robust convolutional neural network for
  retinal blood vessel segmentation}, in: \bibinfo{booktitle}{2020 IEEE
  International Conference on Image Processing (ICIP)},
  \bibinfo{organization}{IEEE}. pp. \bibinfo{pages}{398--402}.
\bibitem[{Liu et~al.(2021a)Liu, Xing, Yang, Kuo, Babu, El~Fakhri, Jenkins and
  Woo}]{liu2021voxelhop}
\bibinfo{author}{Liu, X.}, \bibinfo{author}{Xing, F.}, \bibinfo{author}{Yang,
  C.}, \bibinfo{author}{Kuo, C.C.J.}, \bibinfo{author}{Babu, S.},
  \bibinfo{author}{El~Fakhri, G.}, \bibinfo{author}{Jenkins, T.},
  \bibinfo{author}{Woo, J.}, \bibinfo{year}{2021}a.
\newblock \bibinfo{title}{Voxelhop: Successive subspace learning for als
  disease classification using structural mri}.
\newblock \bibinfo{journal}{IEEE journal of biomedical and health informatics}
  \bibinfo{volume}{26}, \bibinfo{pages}{1128--1139}.
\bibitem[{Liu et~al.(2021b)Liu, Lin, Cao, Hu, Wei, Zhang, Lin and
  Guo}]{liu2021Swin}
\bibinfo{author}{Liu, Z.}, \bibinfo{author}{Lin, Y.}, \bibinfo{author}{Cao,
  Y.}, \bibinfo{author}{Hu, H.}, \bibinfo{author}{Wei, Y.},
  \bibinfo{author}{Zhang, Z.}, \bibinfo{author}{Lin, S.}, \bibinfo{author}{Guo,
  B.}, \bibinfo{year}{2021}b.
\newblock \bibinfo{title}{Swin transformer: Hierarchical vision transformer
  using shifted windows}.
\newblock \bibinfo{journal}{International Conference on Computer Vision (ICCV)}
  .
\bibitem[{Maani et~al.(2016)Maani, Yang, Emery and Kalra}]{maani2016cerebral}
\bibinfo{author}{Maani, R.}, \bibinfo{author}{Yang, Y.H.},
  \bibinfo{author}{Emery, D.}, \bibinfo{author}{Kalra, S.},
  \bibinfo{year}{2016}.
\newblock \bibinfo{title}{Cerebral degeneration in amyotrophic lateral
  sclerosis revealed by 3-dimensional texture analysis}.
\newblock \bibinfo{journal}{Frontiers in neuroscience} \bibinfo{volume}{10},
  \bibinfo{pages}{120}.
\bibitem[{Mok and Chung(2022)}]{mok2022affine}
\bibinfo{author}{Mok, T.C.W.}, \bibinfo{author}{Chung, A.C.S.},
  \bibinfo{year}{2022}.
\newblock \bibinfo{title}{Affine medical image registration with coarse-to-fine
  vision transformer}.
\newblock \bibinfo{journal}{arXiv preprint arXiv:2203.15216v2} .
\bibitem[{Paszke et~al.(2019)Paszke, Gross, Massa, Lerer, Bradbury, Chanan,
  Killeen, Lin, Gimelshein, Antiga et~al.}]{paszke2019pytorch}
\bibinfo{author}{Paszke, A.}, \bibinfo{author}{Gross, S.},
  \bibinfo{author}{Massa, F.}, \bibinfo{author}{Lerer, A.},
  \bibinfo{author}{Bradbury, J.}, \bibinfo{author}{Chanan, G.},
  \bibinfo{author}{Killeen, T.}, \bibinfo{author}{Lin, Z.},
  \bibinfo{author}{Gimelshein, N.}, \bibinfo{author}{Antiga, L.}, et~al.,
  \bibinfo{year}{2019}.
\newblock \bibinfo{title}{Pytorch: An imperative style, high-performance deep
  learning library}.
\newblock \bibinfo{journal}{Advances in neural information processing systems}
  \bibinfo{volume}{32}.
\bibitem[{Pinaya et~al.(2022)Pinaya, Tudosiu, Gray, Rees, Nachev, Ourselin and
  Cardoso}]{pinaya2022unsupervised}
\bibinfo{author}{Pinaya, W.H.}, \bibinfo{author}{Tudosiu, P.D.},
  \bibinfo{author}{Gray, R.}, \bibinfo{author}{Rees, G.},
  \bibinfo{author}{Nachev, P.}, \bibinfo{author}{Ourselin, S.},
  \bibinfo{author}{Cardoso, M.J.}, \bibinfo{year}{2022}.
\newblock \bibinfo{title}{Unsupervised brain imaging 3d anomaly detection and
  segmentation with transformers}.
\newblock \bibinfo{journal}{Medical Image Analysis} \bibinfo{volume}{79},
  \bibinfo{pages}{102475}.
\bibitem[{Playout et~al.(2022)Playout, Duval, Boucher and
  Cheriet}]{playout2022focused}
\bibinfo{author}{Playout, C.}, \bibinfo{author}{Duval, R.},
  \bibinfo{author}{Boucher, M.C.}, \bibinfo{author}{Cheriet, F.},
  \bibinfo{year}{2022}.
\newblock \bibinfo{title}{Focused attention in transformers for interpretable
  classification of retinal images}.
\newblock \bibinfo{journal}{Medical Image Analysis} \bibinfo{volume}{82},
  \bibinfo{pages}{102608}.
\bibitem[{Rao et~al.(2021)Rao, Zhao, Zhu, Lu and Zhou}]{rao2021global}
\bibinfo{author}{Rao, Y.}, \bibinfo{author}{Zhao, W.}, \bibinfo{author}{Zhu,
  Z.}, \bibinfo{author}{Lu, J.}, \bibinfo{author}{Zhou, J.},
  \bibinfo{year}{2021}.
\newblock \bibinfo{title}{Global filter networks for image classification}, in:
  \bibinfo{booktitle}{Advances in Neural Information Processing Systems
  (NeurIPS)}.
\bibitem[{Sadri et~al.(2020)Sadri, Janowczyk, Zhou, Verma, Beig, Antunes,
  Madabhushi, Tiwari and Viswanath}]{sadri2020mrqy}
\bibinfo{author}{Sadri, A.R.}, \bibinfo{author}{Janowczyk, A.},
  \bibinfo{author}{Zhou, R.}, \bibinfo{author}{Verma, R.},
  \bibinfo{author}{Beig, N.}, \bibinfo{author}{Antunes, J.},
  \bibinfo{author}{Madabhushi, A.}, \bibinfo{author}{Tiwari, P.},
  \bibinfo{author}{Viswanath, S.E.}, \bibinfo{year}{2020}.
\newblock \bibinfo{title}{Mrqy—an open-source tool for quality control of mr
  imaging data}.
\newblock \bibinfo{journal}{Medical physics} \bibinfo{volume}{47},
  \bibinfo{pages}{6029--6038}.
\bibitem[{Sage et~al.(2007)Sage, Peeters, G{\"o}rner, Robberecht and
  Sunaert}]{sage2007quantitative}
\bibinfo{author}{Sage, C.A.}, \bibinfo{author}{Peeters, R.R.},
  \bibinfo{author}{G{\"o}rner, A.}, \bibinfo{author}{Robberecht, W.},
  \bibinfo{author}{Sunaert, S.}, \bibinfo{year}{2007}.
\newblock \bibinfo{title}{Quantitative diffusion tensor imaging in amyotrophic
  lateral sclerosis}.
\newblock \bibinfo{journal}{Neuroimage} \bibinfo{volume}{34},
  \bibinfo{pages}{486--499}.
\bibitem[{Stegm{\"u}ller et~al.(2022)Stegm{\"u}ller, Spahr, Bozorgtabar and
  Thiran}]{stegmuller2022scorenet}
\bibinfo{author}{Stegm{\"u}ller, T.}, \bibinfo{author}{Spahr, A.},
  \bibinfo{author}{Bozorgtabar, B.}, \bibinfo{author}{Thiran, J.P.},
  \bibinfo{year}{2022}.
\newblock \bibinfo{title}{Scorenet: Learning non-uniform attention and
  augmentation for transformer-based histopathological image classification}.
\newblock \bibinfo{journal}{arXiv preprint arXiv:2202.07570} .
\bibitem[{Tan and Le(2019)}]{tan2019efficientnet}
\bibinfo{author}{Tan, M.}, \bibinfo{author}{Le, Q.}, \bibinfo{year}{2019}.
\newblock \bibinfo{title}{Efficientnet: Rethinking model scaling for
  convolutional neural networks}, in: \bibinfo{booktitle}{International
  Conference on Machine Learning}, \bibinfo{organization}{PMLR}. pp.
  \bibinfo{pages}{6105--6114}.
\bibitem[{Thome et~al.(2022)Thome, Steinbach, Grosskreutz, Durstewitz and
  Koppe}]{thome2022classification}
\bibinfo{author}{Thome, J.}, \bibinfo{author}{Steinbach, R.},
  \bibinfo{author}{Grosskreutz, J.}, \bibinfo{author}{Durstewitz, D.},
  \bibinfo{author}{Koppe, G.}, \bibinfo{year}{2022}.
\newblock \bibinfo{title}{Classification of amyotrophic lateral sclerosis by
  brain volume, connectivity, and network dynamics}.
\newblock \bibinfo{journal}{Human brain mapping} \bibinfo{volume}{43},
  \bibinfo{pages}{681--699}.
\bibitem[{Touvron et~al.(2021)Touvron, Cord, Douze, Massa, Sablayrolles and
  J{\'e}gou}]{touvron2021training}
\bibinfo{author}{Touvron, H.}, \bibinfo{author}{Cord, M.},
  \bibinfo{author}{Douze, M.}, \bibinfo{author}{Massa, F.},
  \bibinfo{author}{Sablayrolles, A.}, \bibinfo{author}{J{\'e}gou, H.},
  \bibinfo{year}{2021}.
\newblock \bibinfo{title}{Training data-efficient image transformers \&
  distillation through attention}, in: \bibinfo{booktitle}{International
  Conference on Machine Learning}, \bibinfo{organization}{PMLR}. pp.
  \bibinfo{pages}{10347--10357}.
\bibitem[{Valanarasu et~al.(2021)Valanarasu, Oza, Hacihaliloglu and
  Patel}]{valanarasu2021medical}
\bibinfo{author}{Valanarasu, J.M.J.}, \bibinfo{author}{Oza, P.},
  \bibinfo{author}{Hacihaliloglu, I.}, \bibinfo{author}{Patel, V.M.},
  \bibinfo{year}{2021}.
\newblock \bibinfo{title}{Medical transformer: Gated axial-attention for
  medical image segmentation}, in: \bibinfo{booktitle}{International Conference
  on Medical Image Computing and Computer-Assisted Intervention},
  \bibinfo{organization}{Springer}. pp. \bibinfo{pages}{36--46}.
\bibitem[{Vaswani et~al.(2017)Vaswani, Shazeer, Parmar, Uszkoreit, Jones,
  Gomez, Kaiser and Polosukhin}]{vaswani2017attention}
\bibinfo{author}{Vaswani, A.}, \bibinfo{author}{Shazeer, N.},
  \bibinfo{author}{Parmar, N.}, \bibinfo{author}{Uszkoreit, J.},
  \bibinfo{author}{Jones, L.}, \bibinfo{author}{Gomez, A.N.},
  \bibinfo{author}{Kaiser, {\L}.}, \bibinfo{author}{Polosukhin, I.},
  \bibinfo{year}{2017}.
\newblock \bibinfo{title}{Attention is all you need}, in:
  \bibinfo{booktitle}{Advances in neural information processing systems}, pp.
  \bibinfo{pages}{5998--6008}.
\bibitem[{Wang et~al.(2020)Wang, Foxley, Ansorge, Bangerter-Christensen, Chiew,
  Leonte, Menke, Mollink, Pallebage-Gamarallage, Turner
  et~al.}]{wang2020methods}
\bibinfo{author}{Wang, C.}, \bibinfo{author}{Foxley, S.},
  \bibinfo{author}{Ansorge, O.}, \bibinfo{author}{Bangerter-Christensen, S.},
  \bibinfo{author}{Chiew, M.}, \bibinfo{author}{Leonte, A.},
  \bibinfo{author}{Menke, R.A.}, \bibinfo{author}{Mollink, J.},
  \bibinfo{author}{Pallebage-Gamarallage, M.}, \bibinfo{author}{Turner, M.R.},
  et~al., \bibinfo{year}{2020}.
\newblock \bibinfo{title}{Methods for quantitative susceptibility and r2*
  mapping in whole post-mortem brains at 7t applied to amyotrophic lateral
  sclerosis}.
\newblock \bibinfo{journal}{NeuroImage} \bibinfo{volume}{222},
  \bibinfo{pages}{117216}.
\bibitem[{Wen et~al.(2020)Wen, Thibeau-Sutre, Diaz-Melo, Samper-Gonz{\'a}lez,
  Routier, Bottani, Dormont, Durrleman, Burgos, Colliot
  et~al.}]{wen2020convolutional}
\bibinfo{author}{Wen, J.}, \bibinfo{author}{Thibeau-Sutre, E.},
  \bibinfo{author}{Diaz-Melo, M.}, \bibinfo{author}{Samper-Gonz{\'a}lez, J.},
  \bibinfo{author}{Routier, A.}, \bibinfo{author}{Bottani, S.},
  \bibinfo{author}{Dormont, D.}, \bibinfo{author}{Durrleman, S.},
  \bibinfo{author}{Burgos, N.}, \bibinfo{author}{Colliot, O.}, et~al.,
  \bibinfo{year}{2020}.
\newblock \bibinfo{title}{Convolutional neural networks for classification of
  alzheimer's disease: Overview and reproducible evaluation}.
\newblock \bibinfo{journal}{Medical image analysis} \bibinfo{volume}{63},
  \bibinfo{pages}{101694}.
\bibitem[{Yagis et~al.(2021)Yagis, Atnafu, Garc{\'\i}a Seco~de Herrera, Marzi,
  Scheda, Giannelli, Tessa, Citi and Diciotti}]{yagis2021effect}
\bibinfo{author}{Yagis, E.}, \bibinfo{author}{Atnafu, S.W.},
  \bibinfo{author}{Garc{\'\i}a Seco~de Herrera, A.}, \bibinfo{author}{Marzi,
  C.}, \bibinfo{author}{Scheda, R.}, \bibinfo{author}{Giannelli, M.},
  \bibinfo{author}{Tessa, C.}, \bibinfo{author}{Citi, L.},
  \bibinfo{author}{Diciotti, S.}, \bibinfo{year}{2021}.
\newblock \bibinfo{title}{Effect of data leakage in brain mri classification
  using 2d convolutional neural networks}.
\newblock \bibinfo{journal}{Scientific reports} \bibinfo{volume}{11},
  \bibinfo{pages}{1--13}.
\bibitem[{Yan et~al.(2020)Yan, Huang, Xia, Gu, Yan, Wang and Tao}]{yan2020mri}
\bibinfo{author}{Yan, W.}, \bibinfo{author}{Huang, L.}, \bibinfo{author}{Xia,
  L.}, \bibinfo{author}{Gu, S.}, \bibinfo{author}{Yan, F.},
  \bibinfo{author}{Wang, Y.}, \bibinfo{author}{Tao, Q.}, \bibinfo{year}{2020}.
\newblock \bibinfo{title}{Mri manufacturer shift and adaptation: increasing the
  generalizability of deep learning segmentation for mr images acquired with
  different scanners}.
\newblock \bibinfo{journal}{Radiology: Artificial Intelligence}
  \bibinfo{volume}{2}.
\bibitem[{Zhang et~al.(2018)Zhang, Zhou, Lin and Sun}]{zhang2018shufflenet}
\bibinfo{author}{Zhang, X.}, \bibinfo{author}{Zhou, X.}, \bibinfo{author}{Lin,
  M.}, \bibinfo{author}{Sun, J.}, \bibinfo{year}{2018}.
\newblock \bibinfo{title}{Shufflenet: An extremely efficient convolutional
  neural network for mobile devices}, in: \bibinfo{booktitle}{Proceedings of
  the IEEE conference on computer vision and pattern recognition}, pp.
  \bibinfo{pages}{6848--6856}.
\bibitem[{Zhang et~al.(2021)Zhang, Liu and Hu}]{zhang2021transfuse}
\bibinfo{author}{Zhang, Y.}, \bibinfo{author}{Liu, H.}, \bibinfo{author}{Hu,
  Q.}, \bibinfo{year}{2021}.
\newblock \bibinfo{title}{Transfuse: Fusing transformers and cnns for medical
  image segmentation}, in: \bibinfo{booktitle}{International Conference on
  Medical Image Computing and Computer-Assisted Intervention},
  \bibinfo{organization}{Springer}. pp. \bibinfo{pages}{14--24}.

\end{thebibliography}

\end{document}